\documentclass[a4paper,fleqn]{cas-sc}

\usepackage[numbers]{natbib}
\usepackage{epsfig}
\usepackage{float}
\usepackage{graphicx}
\usepackage{amssymb}
\usepackage{amsthm}
\usepackage{mathtools}
\usepackage{bm}
\usepackage{soul}
\usepackage{subcaption}
\usepackage{caption}
\usepackage{placeins}
\usepackage[outdir=./figures/]{epstopdf}
\usepackage{listings}
\usepackage{tcolorbox}
\tcbuselibrary{listings}
\definecolor{lgray}{HTML}{D4D7D9}

\lstdefinestyle{customc}{
  belowcaptionskip=1\baselineskip,
  xleftmargin=\parindent,
  language=C,
  showstringspaces=false,
  basicstyle=\footnotesize,
  numbers=left,
  keywordstyle=\bfseries\color{green!40!black},
  commentstyle=\color{purple!30!black},
  identifierstyle=\color{blue},
  stringstyle=\color{orange},
  aboveskip=0pt,
  belowskip=0pt
}

\newtcblisting{mylisting}{
      arc=0pt,
      top=0mm,
      bottom=0mm,
      left=0mm,
      right=0mm,
      boxrule=0pt,
      colback =lgray,
      listing only,
      listing options={style=customc},
      hbox
}

\newcommand{\reg}{\textsuperscript{\textregistered}~}
\newcommand{\ofx}{\left( \bm{x} \right)}

\newcommand{\oft}{\left( t \right)}
\newcommand{\ofxt}{\left( \bm{x}, t \right)}

\renewcommand{\div}{\boldsymbol{\nabla} \cdot}
\newcommand{\grad}{\boldsymbol{\nabla} }
\newcommand{\ddt}[1]{\frac{\partial #1}{\partial t} }

\newcommand{\Mdot}{\dot{\mathcal{M}}}

\usepackage{mymacros_10Sep19}

    \usepackage{framed} 
    \usepackage{multicol} 
     
    \usepackage{nomencl} 
    \makenomenclature
    \renewcommand*\nompreamble{\begin{multicols}{2}}
    \renewcommand*\nompostamble{\end{multicols}}

\newcommand{\diff}{\mathcal{D}}

\newcommand{\mc}{c^{\prime}}
\newcommand{\mU}{\bm{u}^{\prime}}
\newcommand{\mD}{\mathcal{D}^{\prime}}
\newcommand{\tc}{\text{c}}
\newcommand{\tf}{\text{f}}
\usepackage{ctable} 

\newcommand{\OF}{\textsc{OpenFOAM}\textsuperscript{\textregistered}}

\graphicspath{{figures/}}

\newcommand{\citfull}[1]{\citeauthor{#1} \citep{#1}}
\usepackage{cleveref}
\begin{document}
\let\WriteBookmarks\relax
\def\floatpagepagefraction{1}
\def\textpagefraction{.001}
\shorttitle{Heterogeneous Multi-Rate mass transfer models in \textsc{OpenFOAM}\reg}
\shortauthors{F. Municchi et~al.}

\title [mode = title]{Heterogeneous Multi-Rate mass transfer models in \textsc{OpenFOAM}\reg}                      



\author[1]{Federico Municchi}


\address[1]{School of Mathematical Sciences, University of Nottingham, NG7 2RD, UK}
\address[2]{Institute of Environmental Assessment and Water Research (IDÆA-CSIC), Barcelona, Spain}
\author[1]{Nicodemo di Pasquale}
\author[2]{Marco Dentz}
\author[1]{Matteo Icardi}
\cormark[1]






\begin{abstract}
We implement the Multi-Rate Mass Transfer (MRMT) model for mobile-immobile transport in porous media \citep{Haggerty1995,gmrmt2019} within the open-source finite volume library \textsc{OpenFOAM}\reg \citep{Foundation2014}. Unlike other codes available in the literature \citep{Geiger2011,Silva2009}, we propose an implementation that can be applied to complex three-dimensional geometries and highly heterogeneous fields, where the parameters of the MRMT can arbitrarily vary in space. Furthermore, being built over the widely diffused \textsc{OpenFOAM}\reg library, it can be easily extended and included in other models, and run in parallel.%
We briefly describe the structure of the \emph{multiContinuumModels} library that includes the formulation of the MRMT based on the works of \citfull{Haggerty1995} and \citfull{Municchi2020}.
The  implementation is verified against benchmark solutions and tested on two- and three-dimensional random  permeability fields. The role of various physical and numerical parameters, including the transfer rates, the heterogeneities, and the number of terms in the MRMT expansions, is investigated. 
Finally, we illustrate the significant role played by heterogeneity in the mass transfer when permeability and porosity are represented using Gaussian random fields.     
\end{abstract}


\begin{highlights}
\item Implementation of the Multi-Rate mass transfer model in {OpenFOAM}\reg
\item The model allows heterogeneous fields and arbitrary geometry
\item Comparison against analytical solutions   
\item {OpenFOAM}\reg based workflow for geological simulations.  
\end{highlights}


\maketitle


\nomenclature{\mbox{$c_m$}}{Concentration in the mobile region}
\nomenclature{\mbox{$c_i$}}{Concentration in the immobile region $i$}
\nomenclature{\mbox{$c_{im}$}}{Average concentration in the immobile region }
\nomenclature{\mbox{$\mathcal{D}_m$}}{Diffusion coefficient for the mobile region}
\nomenclature{\mbox{$\mathcal{D}_i$}}{Diffusion coefficient for immobile region $i$}
\nomenclature{\mbox{$\beta_m$}}{Capacity in the mobile region}
\nomenclature{\mbox{$\beta_i$}}{Capacity in the immobile region $i$}
\nomenclature{\mbox{$\bm{u}$}}{Velocity field}
\nomenclature{\mbox{$\lambda_{ik}$}}{Eigenvalue for the immobile region $i$ mode $k$}
\nomenclature{\mbox{$\alpha_{ik}$}}{Dimensionless eigenvalue for the immobile region $i$ mode $k$}
\nomenclature{\mbox{$\omega_{i}$}}{Characteristic transfer frequency for immobile region $i$} 
\nomenclature{\mbox{$N_i$}}{Number of immobile regions}
\nomenclature{\mbox{$\overline{(*)}$}}{Volume averaging operator}
\nomenclature{\mbox{$(*)^{\prime}$}}{Microscopic quantity}
\nomenclature{\mbox{$V$}}{Reference volume}
\nomenclature{\mbox{$V_i$}}{Volume of immobile region $i$}
\nomenclature{\mbox{$V_m$}}{Volume of the mobile region}
\nomenclature{\mbox{$\Omega_m$}}{Mobile region}
\nomenclature{\mbox{$\Omega_i$}}{Immobile region $i$}
\nomenclature{\mbox{$\dot{\mathcal{M}}_i$}}{Inter-region mass exchange for region $i$}
\nomenclature{\mbox{$\mathbf{J}_m$}}{Effective flux in the mobile region}
\nomenclature{\mbox{$(*)_{\tc}$}}{Quantity evaluated at the cell centre}
\nomenclature{\mbox{$(*)_{\tf}$}}{Quantity evaluated at the face centre}
\nomenclature{\mbox{$\bm{n}_{\tc,\tf}$}}{Normal to face $\tf$ from cell $\tc$}
\nomenclature{\mbox{$S_{\tc,\tf}$}}{Surface of face $\tf$ from cell $\tc$}
\nomenclature{\mbox{$N_{i}$}}{Number of immobile regions}
\nomenclature{\mbox{$N_{\tc}$}}{Number of cells}
\nomenclature{\mbox{$N_{\tc,tf}$}}{Number of faces per cell $\tc$}
\nomenclature{\mbox{$V_{\tc}$}}{Cell volume}
\nomenclature{\mbox{$Co$}}{Courant number}
\nomenclature{\mbox{$\beta_{\text{tr}}$}}{Truncation $\beta$}
\nomenclature{\mbox{$\zeta_{k}$}}{$k$-th zero of the Bessel function of the first kind $J_0$}
\nomenclature{\mbox{$c_{eq}$}}{Equilibrium concentration}
\nomenclature{\mbox{$\hat{c}_{m}$}}{Normalised concentration in the mobile region}
\nomenclature{\mbox{$M$}}{Number of terms retained in the expansion}
\nomenclature{\mbox{$P$}}{Pressure field}
\nomenclature{\mbox{$K$}}{Permeability tensor field}
\nomenclature{\mbox{$k$}}{Magnitude of the permeability field}
\nomenclature{\mbox{$a$}}{Conversion coefficient}
\nomenclature{\mbox{$BT$}}{Breakthrough}
\nomenclature{\mbox{$S_{out}$}}{Boundary corresponding to the domain outlet}
\nomenclature{\mbox{CCS}}{Carbon Capture and Storage}
\nomenclature{\mbox{$L_{i}$}}{Characteristic length of the immobile region}

\noindent
{\bf Program summary}

\begin{small}
\noindent
{\em Program Title:} mrmtFoam                                      \\
{\em Developer's repository link:}  \url{https://github.com/multiform-UoN/mrmtFOAM}                                    \\
{\em Code Ocean capsule:}      
\\
{\em Licensing provisions:} GPL 3.0
\\
{\em Nature of problem:} Large scale dynamics of heat and mass transfer in heterogeneous media where one mobile region coexists with multiple immobile regions.
   \\
{\em Solution method:} The multi-rate mass transfer model is employed to described pre-asymptotic (i.e., non equilibrium ) transfer between regions. This method is implemented using the opensource finite volume library OpenFOAM\textsuperscript{\textregistered} 
   \\
{\em References:} \citfull{federico_municchi_2020_3938868}
\end{small}

\nomenclature{\mbox{MRMT}}{Multi-Rate mass transfer}

\begin{table*}
\begin{framed}
\printnomenclature
\end{framed}
\end{table*}



\section{Introduction}


\noindent Interests in porous media dates back to the middle of nineteenth century with the study of Darcy describing the law that bears his name \citep{Whitaker1986}. The reasons for such a long lasting interest reside in the fact that porous media are present in a wide range of systems and applications, both as natural \cite{Zou2017} or industrial synthetic media.
Also, environmental applications as risk and safety assessment of groundwater contamination \citep{Tang1981,Sudicky1982}, reservoir storage, geothermal extraction, geological disposal of radioactive waste \citep{Neretnieks1980} and carbon dioxide require the study of fluid flow and solute transport in heterogeneous porous and fractured media.

One of the key characteristic of these systems is their heterogeneity, which results in non-equilibrium and memory effects \citep{Dentz2018,Crevacore2016a}. A porous medium can be described as a matrix of solid material in which a fluid phase moves. Usually, the region occupied by the fluid phase (mono or multi-component) is called \textit{mobile} region whereas the remaining region, occupied by the matrix, is the \textit{immobile} region. It is often assumed that the dominant transport process in the immobile regions is diffusion \citep{Municchi2020}, while the mobile region can exchange mass and energy with the immobile region. Figure \ref{fig:mrmt_illustration} depicts a typical domain composed of a mobile region and several immobile regions, similar to those often found in subsurface flow applications. Furthermore, such classification into mobile and immobile regions can be applied to fluid and fillers in packed beds or even to circulation and recirculation zones in fluid flows  \citep{Zhou2019}.

The complexity of flow fields into a highly heterogeneous medium (such as a porous medium) modify the transport behaviour of solutes within the fluid. In these kinds of systems non-Fickian transport behaviour is observed \citep{Neuman2009,Dentz2011b}. The nature of the non-Fickian transport was extensively studied in literature \cite{Berkowitz2008,Berkowitz2009,Neuman2009}, and its origin was found in the broad spectrum of transition times intrinsic to heterogeneous media \citep{Berkowitz2006}.

Generally, immobile regions do not just introduce heterogeneities leading to non-Fickian dispersion in the porous medium, but they also act as storage (of heat or mass), leading to the breaking of time locality. As a result, the dynamics is non-local in time and therefore, it depends on the history of the system. Several methods were devised to mathematically describe such phenomenon in porous media.  Among the spectrum of methodologies, we want to recall:
\begin{itemize}
    \item The Dual-Porosity formulation \citep{Moench1984}, where the medium is modelled as two overlapping continua (mobile and immobile);
    \item The Integro-Differential formulation \citep{Herrera1973,Herrera1977,Carrera1998}, which uses a mix of convolutions of state variables and memory functions to model  non-local mass transfer;
    \item The Continuous Time Random Walk \citep{Berkowitz1998, Dentz2004,Berkowitz2006}, which assumes a random walk in time and space for the movement of solute particles in a heterogeneous medium.
\end{itemize}

One of the formulations which received significant attention in the past decades is the Multi-Rate Mass Transfer (MRMT) Model \citep{Roth1993,Haggerty1995,Haggerty2000,Wang2005,Gouze2008,Benson2009} which can be seen as a generalisation of the dual-porosity model. In the MRMT one performs a spectral decomposition of the diffusion operator in each immobile region, leading to infinite series of response terms representing the non-local transfer \citep{Municchi2020}. Thus, mass transfer between mobile and immobile regions is linear and can be represented by a suite of first-order processes or diffusive mass transfer processes. The interest in the MRMT model also stems from being mathematically equivalent to the other models described earlier \citep{Dentz2003,Silva2009}, while maintaining  the important property of localisation. This means that the state of the heterogeneous systems can be described, at any instant of time, locally and without the need to define global quantities, which usually allow less flexibility in the description of these complex systems. Notice that in the MRMT model, a spectral decomposition is only performed in the immobile regions, which are often assumed to be simply connected and not complicated in shape. Therefore, the eigenfunctions can be easily obtained analytically in a large number of practical cases. However, the MRMT model does not apply to the mobile region, where the fluid is flowing. Therefore, macroscopic quantities like the effective fluxes must be obtained using other methods (for example, classical volume averaging \citep{Whitaker1999a} or two-scale asymptotics \citep{Municchi2020a}). But this does not play a role in the numerical implementation of the MRMT model, which governs the inter-region transfer process. 

A wide range problems can be addressed by the MRMT methodology, a fact that makes this framework extremely flexible and useful.  
An important environmental problem we are currently facing, is the high concentration of CO$_2$ in the atmosphere, due to anthropogenic contribution (mainly energy production). One of the most promising strategies devised to reduce carbon emissions is the Carbon Capture and Storage (CCS) through gas hydrate and crystallisation \citep{Song2013}. This process separates the CO$_2$ from fuel gas, by sequestration in gas hydrate crystals (mainly water). It was observed this process can be improved using porous media, thanks to their much high gas/water contact area \citep{Adeyemo2010}. Therefore, a reliable model for transport properties in such systems becomes essential. 

Another environmental application is dispersion of contaminants in aquifer and groundwater remediation, where the dynamic of diffusion of chemical compounds into the ground is studied. One of the  way to perform \textit{in situ} remediation which has recently gained significant attention, involves the injection of a reactive suspension of engineered nanoparticles to degrade transform, or immobilise the pollutants\citep{Georgi2015,Yan2013}. Both the dynamics of the dispersion of the pollutant, and the transport of these nanoparticles  \citep{Liu2011} are dominated by the exchanging of mass or energy with a set of impermeable inclusions, which could be modelled through MRMT. 

A formulation of the numerical implementation of the MRMT model was presented in \citfull{Silva2009}. This approach is capable of describing a wide range of non-equilibrium phenomena by using a model which is local in time. Variables referring to immobile regions are solved as explicit functions, avoiding the need of a discretisation of these regions. By assuming a functional form for the the concentration in the mobile region during each time increment (in the  \citep{Silva2009} it was assumed a linear behaviour), it is possible to explicitly integrate the first-order linear differential equations referring to the immobile regions. Therefore, in this formulation the explicit contribution of each immobile region can be included in the discretised (in time and and space) equations describing the evolution of the concentration in the mobile region.

In this work we propose a novel numerical implementation of the MRMT model, based on the generalisation proposed in an earlier work of some of the authors \citep{Municchi2020}. This numerical implementation is written as a new library within the C$++$ opensource finite volume library \textsc{OpenFOAM}\reg \citep{Openfoam2019}. Our choice of \textsc{OpenFOAM}\reg was based on the fact that this code has a wide diffusion across industry and academia alike, and has a solid and active community of users.

This paper is structured as follows: we first present the relevant equations and hypothesis for the generalised MRMT model by following the derivation reported in \citep{Municchi2020}. We then proceed to show how these equations can be implemented in \textsc{OpenFOAM}\reg by describing the structure of the new library we are presenting and we report some examples of application of our library in a number of cases. We then proceed to draw some conclusions and outlooks on possible further uses of this library.

\section{Mathematical formulation of the MRMT model}




\noindent We present here the theoretical background of the MRMT model and we refer to \citep{Municchi2020} for a complete derivation of the relevant equations. However, notice that the notation we employ is slightly different from that used in \citep{Municchi2020}, where the main focus was on the derivation of macroscopic equations from the microscale dynamics. The present work focuses on the macroscale exclusively and therefore, we will use primed symbols to indicate microscopic quantities and un-primed symbols to indicate macroscopic averaged quantities.  

One point we want to emphasize here is that, while we are using the symbol $c$ for the main quantities to be intended as a concentration of some chemical species in the domanin, the equations can be easily adapted to any other scalar quantity (such as the temperature). Our use of this terminology comes form the fact that this model comes from geological applications and we wanted to be consisted with the terminology used in this field.

\begin{figure}
    \centering
    \includegraphics[width=.8\linewidth]{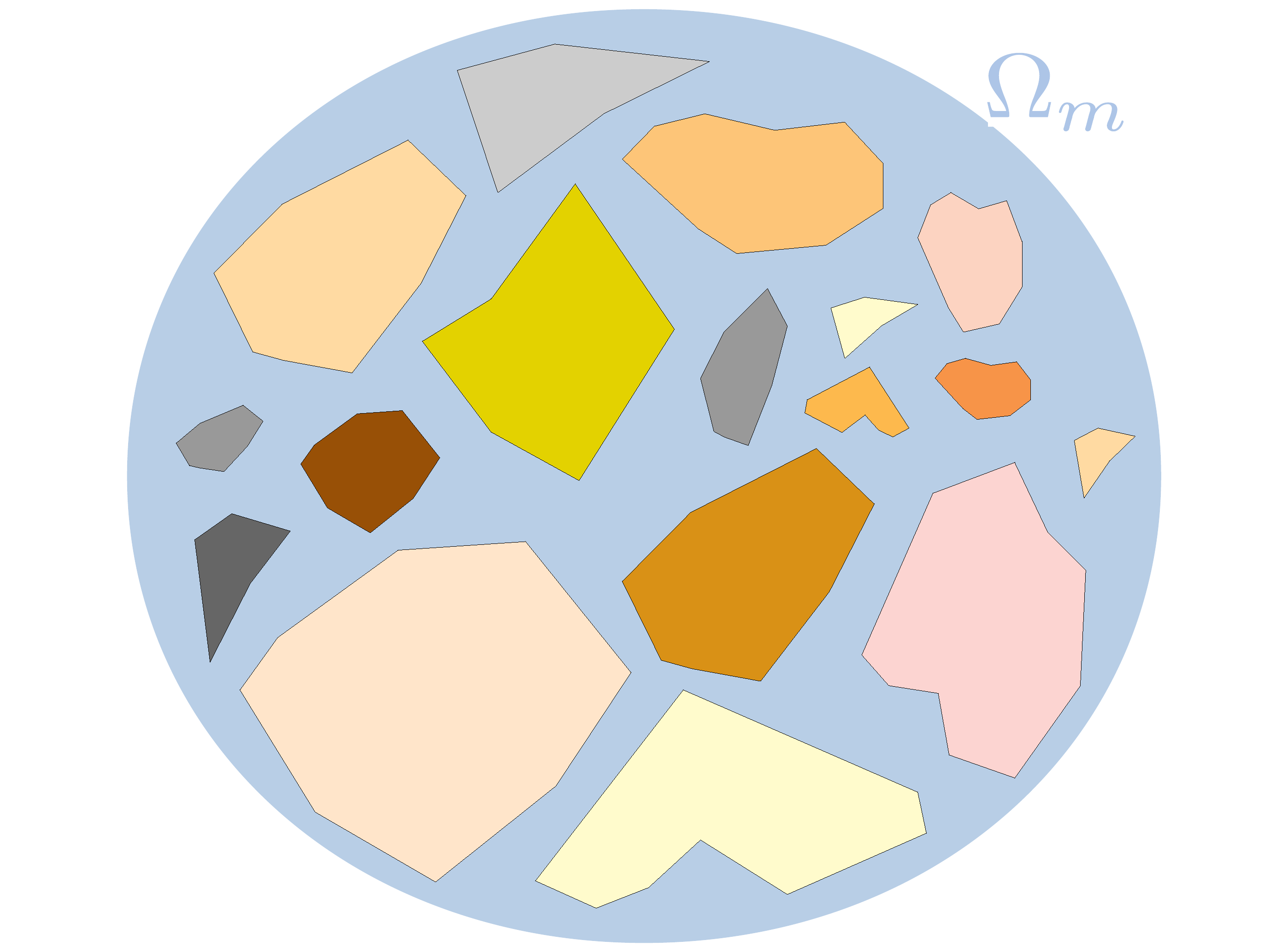}
    \caption{Representation of a typical heterogeneous medium with a mobile region (light blue) and several immobile regions (colours) with different shape and composition. Fluid is flowing in the mobile region only, while mass (or heat) can diffuse within the immobile regions.}
    \label{fig:mrmt_illustration}
\end{figure}

Let us consider a heterogeneous domain $\Omega$ composed of a \textit{mobile} region, $\Omega_m$, and a number $N$ of \textit{immobile} regions, $\Omega_i$, with $i=1,\ldots,N$ such that $\Omega=\Omega_m \bigcup_{i=1}^{N}\Omega_i$ (see Figure \ref{fig:mrmt_illustration}).

We assume the transport in the mobile region can be modelled by the advection-diffusion equation:
\begin{equation}
    \label{eq::AD_m}
    \ddt{\mc_m} + \div \left( \mU \mc_m - \mD_m \grad \mc_m \right) = 0, \quad \bm{x} \in \Omega_m\,.
\end{equation}
where $\mc_m$ is a scalar field (which will be referred to as "concentration" for simplicity, even if it could be a temperature field as well), $\mD_m$ is the diffusion coefficient in the mobile region and $\mU$ is the velocity field. 
Furthermore, we consider $N_i$ diffusion equations, one for each immobile region describing the concentrations $\mc_i(\bfx,t)$ in the $i$-th immobile region:
\begin{equation}
    \label{eq::D_imm}
    \ddt{\mc_i} = \mD_i \nabla^2 \mc_i, \quad \bm{x} \in \Omega_i, \quad i=1,\dots,N_i.
\end{equation}
where $\mD_i$ is the diffusion coefficient in the $i$-th immobile region. Note that in previous equations we dropped the spatial and temporal dependence from $\mc_m$ and $\mc_i$ to simplify the notation. 
The equations describing the evolution of the concentration field in the mobile and the immobile regions must be coupled with the proper boundary conditions between the two kind of regions. Specifically, we impose continuity of fields and fluxes through the interfaces:
\begin{equation}
    \label{eq::bc}
    \mc_i = \mc_m, \quad \mD_i \frac{\partial \mc_i}{\partial n}  = \mD_m \frac{\partial \mc_m}{\partial n}, \quad \bm{x} \in \partial \Omega_i.    
\end{equation}
An important implication of such boundary condition is that each immobile region is connected only to the mobile region, i.e. the immobile regions are not connected. Note that we consider here linear transport of a passive tracer. It is straightforward to account for linear equilibrium sorption in this modeling framework through retardation factors for the mobile and immobile domains as detailed in \cite{Haggerty1995}. Chemical reactions in the mobile and immobile regions can be accounted for by the addition of source and sink terms in the conservation equations for the mobile and immobile species~\cite[e.g.,][]{Lichtner:Kang,Liu:2008,Donado:et:al:2009,Dentz2011,Orgogozo2013}. Also non-linear relations between the scalar concentrations in the mobile and immobile zones can be modeled in this framework as done by \cite{Tecklenburg2013} in the context of two-phase flow in fractured media. 

The next step requires to smooth the concentration field by applying a suitable spatial filtering. In \citep{Municchi2020} two filters were defined: the volume filter over $\Omega$ and the Favre filter acting on the volume of the mobile region $\Omega_m$. We will call $\overline{c}$ the volume averaged concentration field and $c$ (instead of $\mc$) the Favre averaged concentration field. The relation between the two is \citep{Municchi2020}:
\begin{equation}
    c_m = \frac{1}{V_m}\int \limits_{\Omega_m} \mc_m dV\,, \quad
    c_i = \frac{1}{V_i}\int \limits_{\Omega_i} \mc_i dV \,,
    \label{eq:c-Favre}
\end{equation}
\begin{equation}
    \overline{c}_m = \beta_m c_m \, , \quad \overline{c}_i = \beta_i c_i
    \label{eq:ct-cm}
\end{equation}
which allows to write the concentration field in terms of a quantity (the Favre filtered concentration field) specific only to the mobile region. In the previous equation we also introduced the capacity of the mobile region, $\beta_m$, defined as the ratio between the volume of the region $\Omega$, $V$, and the volume of the mobile region, $V_m$, so that $\beta_m=V/V_m$. Similarly, $\beta_i=V/V_i$, being $V_i$ the volume occupied by the $i$-th immobile region.

By applying the volume filtering to \cref{eq::AD_m} and using \cref{eq::bc} and \cref{eq:ct-cm} along with the Gauss-Green theorem, we obtain the following equation for the filtered quantities:
\begin{align}
    \label{eq::mc_1}
  \beta_m \ddt{c_m} + \sum \limits_{i=1}^{N_i}\Mdot_i\oft =
     - \div \bfJ_m\,,
\end{align}
where we introduced the average inter-region mass exchange rate for region $i$, $\Mdot_i\oft$, defined as:
\begin{equation}\label{eq:M}
\Mdot_i\oft = \frac{1}{V} \int \limits_{\partial \Omega_i} \mD_i \frac{\partial \mc_i}{\partial n} \text{d}S 
\end{equation}
and the total average flux in the mobile region, $\bfJ_m$. This last quantity  can be interpreted as an effective flux in the mobile region, $\bfJ_{m,\text{eff}}$, defined as \citep{Municchi2020,Municchi2020a}:
\begin{equation}
    \label{eq::Jeff}
    {\bf{J}}_m = \bm{u} c_m - \diff_{m} \cdot \grad c_m
\end{equation}
where $\bm{u}$ and $\diff_{m}$ are the effective (i.e., macroscopic) velocity and the effective diffusivity (generally a tensor, hence the dot product in equation \ref{eq::Jeff}), which include the contribution of dispersion phenomena. Such quantities can be evaluated employing volume averaging \citep{Whitaker1999a} or homogenisation theory \citep{Municchi2020a,Auriault1995} and in the following, we will assume they are known at each instant of time.
By aplying the Favre average to the concentration in the immobile region,  we obtain from \cref{eq::AD_m}:
\begin{equation}
     \ddt{c_i} = \frac{\Mdot_i\oft }{\beta_i} \, .
\end{equation}
Finally, the multicontinuum equation for the concentration field in the mobile region can be written as:
\begin{equation}
      \beta_m \ddt{c_m} + \sum \limits_{i=1}^{N_i} \beta_i \ddt{c_i} =
     - \div \bfJ_m\,.
\end{equation}
In the latter equation we have included the boundary conditions of the equations at the microscale (see \cref{eq::bc}) as source terms, one for each immobile region. 

A key point of the MRMT formulation, is that the concentration in each immobile region is expressed as linear combination of the eigenfunctions of the diffusion operator. Therefore, the average concentration in each immobile region can be uniquely decomposed as:
\begin{equation}
    \label{eq::c_ik}
    c_i = \sum \limits_{k=0}^{\infty}  \beta_{ik} c_{ik} \, ,
\end{equation}
where $c_{ik}$ is the (unknown and time dependent) coefficient corresponding to the $k$-th eigenfunction, and $\beta_{ik}$ plays the role of a capacity relative to the $k$-th mode.

After performing a number of manipulations involving spectral analysis of the diffusion operator in the immobile regions \cite{Municchi2020}, one can recover the classic formulation of \citfull{Haggerty1995}:
\begin{equation}
    \label{eq::mrmt_HG}
    \begin{dcases}
    &\beta_m \ddt{c_m} +  \sum \limits_{i=1}^{N_i} \beta_i \sum \limits_{k=1}^{\infty} \beta_{ik}\ddt{ c_{ik}}= - \div \mathbf{J}_{m}   \\
    &    \ddt{c_{ik}} = \lambda_{ik}\left(c_{ik} - c_m \right),  {\begin{array}{*{20}c}
  i=1,\ldots,N_i   \\
 k=1,\ldots,\infty  \\    
 \end{array} } 
    \end{dcases}
\end{equation}
where $\lambda_{ik}$ is the eigenvalue corresponding to the $k$-th eigenfunction, which can be written in dimensionless form as:
\begin{equation}\label{eq:dimless}
    \lambda_{ik} = \alpha_{ik}\omega_i \, ,
\end{equation}
where $\omega_i=\frac{\mD_i}{L_i}$ is specific for each immobile region, with $L_i$ the characteristic length of the $i$-th immobile region. Notice that the values of $\alpha_{ik}$ can be computed for any geometrical configuration of the immobile region $i$ following the approach described in \citfull{Municchi2020}. 

\section{Numerical implementation in \textsc{OpenFOAM}\reg}

\subsection{Discretisation method}

The governing equations \ref{eq:dimless} are discretised by means of the Finite Volume Method (FVM), which consists in integrating the governing equations over a set of control volumes (cells) on a numerical mesh. Furthermore, one assumes that the cells are sufficiently small that all fields can be assumed to vary linearly within each cell \cite{Moukalled2016}. Therefore, system \eqref{eq::mrmt_HG} is written in the finite volume formulation:
\begin{equation}
    \label{eq::mrmt_fvm}
    \begin{dcases}
    &\int \limits_{V_{\tc}} \left[\beta_m \ddt{c_m} +  \sum \limits_{i=1}^{N_i} \sum \limits_{k=1}^{\infty} \beta_{ik}\ddt{ c_{ik}} \right] \text{d}V= \sum \limits_{\tf=0}^{N_{\tc,\tf}}
    \oint_{S_{\tc,\tf}}\left( \mathcal{D}_{m} \cdot \grad c_m- \bm{u}c_m \right)\cdot \bm{n}_{\tc,\tf} \text{d}S \\
    &    \int \limits_{V_{\tc}} \left[\ddt{c_{ik}} - \lambda_{ik}\left(c_{ik} - c_m \right)\right] \text{d}V = 0,  {\begin{array}{*{20}c}
  i=1,\ldots,N_i   \\
 k=1,\ldots,\infty  \\   
 \tc = 0,\ldots,N_{\tc}
 \end{array} } 
    \end{dcases}
\end{equation}
where the index $\tc$ runs from $0$ to the number of cells in the mesh $N_{\tc}$, while
the index $\tf$ runs from $0$ to the number of faces $N_{\tc,\tf}$ belonging to cell $\tc$.
Furthermore, $V_{\tc}$ is the volume of cell $\tc$ while $S_{\tc,\tf}$ and $\bm{n}_{\tc,\tf}$   are respectively the surface and surface normal of face $\tf$ in cell $\tc$.

Using the assumption of linearity within each cell and using the subscripts $e$ and $f$ to indicate fields evaluated at the cell and face center respectively, one obtains the discretised system:
\begin{equation}
    \label{eq::mrmt_disc}
    \begin{dcases}
    &\beta_{m,\tc} \ddt{c_{m,\tc}} +  \sum \limits_{i=1}^{N_i} \sum \limits_{k=1}^{\infty} \beta_{ik,\tc}\ddt{ c_{ik,\tc}} = \frac{1}{V_{\tc}} \sum\limits_{f=0}^{N_{\tc,\tf}}
    \left( \mathcal{D}_{m,\tf} \cdot (\grad c_m)_{\tf}- \bm{u}_{\tf} c_{m,\tf} \right)\cdot \bm{n}_{\tc,\tf} S_{\tc,\tf}  \\
    &    \ddt{c_{ik,\tc}} - \lambda_{ik,\tc}\left(c_{ik,\tc} - c_{m,\tc} \right) = 0,  {\begin{array}{*{20}c}
  i=1,\ldots,N_i   \\
 k=1,\ldots,\infty  \\   
 \tc = 0,\ldots,N_\tc
 \end{array} } 
    \end{dcases}
\end{equation}
where the spectral expansion is then truncated such that $k=1\dots M$ where M is the number of terms in the expansion to be  retained.

The FVM requires the definition of appropriate discretisation operators to express face-based variables like $(\nabla c_m)_{\tf}$ and $c_{m,\tf}$ as cell-based variables (e.g., $c_{m,\tc}$). A number of such discretisation methods (e.g., Gauss integration based linear and upwind schemes) are available in \textsc{OpenFOAM}\textsuperscript{\textregistered} \cite{boccardo2020computational,Foundation2014} and they will not be discussed here since their effectiveness depends on the specific problem to solve. Unlike previous works \cite{Silva2009}, the equations for the immobile regions are discretised in time and solved separately and iteratively rather than being included in the equation for the mobile region. While this leads to larger memory requirements and requires a more complex software architecture \cite{Silva2009}, it also results in a more flexible code. Most importantly, it allows to include more complicated physical models in the future and to couple the equations in an implicit manner through multiple iterations.

The \emph{multiContinuumModels} library we are discussing here is publicly available at \citep{federico_municchi_2020_3938868}.

\subsection{Library structure}

The \emph{multiContinuumModels} library \citep{federico_municchi_2020_3938868} follows a flexible object oriented structure. This allows to easily implement new functionalities with relative ease.
A base abstract class named \emph{multiContinuumModel} store references to concentration $c_m$ and capacity $\beta_m$ in the mobile region, and provides public functions for calculating the source term $\Mdot$ as well as for updating the model describing the immobile regions. These are the only functions that need to be called inside an \textsc{OpenFOAM}\textsuperscript{\textregistered} application to make use of this library.

The \emph{multiContinuumModel} class is base for the \emph{multiRateMassTransfer} class, which holds a list of pointers to \emph{immobileRegion} objects (representing the transfer models for each immobile region) and stores the total concentration in the immobile regions $\sum_{i}c_i$. This class also implements the function that returns the overall source term  $\Mdot$.

Classes derived from the \emph{immobileRegion} abstract class are at the core of the multi-rate model, since they implement different kind of transfer models based on the geometrical and physical properties of the medium. Each immobile region holds a list of pointers to fields representing the concentration $c_{ik}$ corresponding to each term in the multi-rate series. The length of such array is given by the \emph{nOfTerms} label, and is read at the beginning of the simulation. Furthermore, this class stores the concentration in the immobile region $c_i$, the relative capacity $\beta_i$, and the frequency $\omega_i$. These are all fields, and can be defined by the user as uniform or non-uniform (e.g., non-uniform initial condition on $c_i$, spatially varying capacity $\beta_i$ and medium properties $\omega_i$). This class also implements a function that solves the system of ODEs for the multi-rate terms. The multi-rate coefficients $\alpha_{ik}$ and $\beta_{ik}$ are computed in derived classes, specialised for spheres, layers, cylinders, and first order regions. The user can easily develop new derived classes given for specific geometries and diffusion  processes, by defining the multi-rate parameters.

\subsection{\emph{multiRateScalarTransportFoam}}
\label{S::multiRateScalarTransportFoam}
 
 The library includes an application for solving the scalar transport equations with the multi-rate mass transfer model. Such application, named \emph{multiRateScalarTransportFoam}, solves the system of equations \ref{eq::mrmt_disc}, corresponding to the multi-rate model of \citfull{Haggerty1995}. The solver is based on the standard \emph{scalarTransportFoam} available in native \textsc{OpenFOAM}\textsuperscript{\textregistered} and employs a special \emph{multiContinuumControl} object (derived from the PIMPLE algorithm in \textsc{OpenFOAM}\textsuperscript{\textregistered}) that wraps the \emph{multiContinuumModel} library and checks for convergence.
 
\begin{figure*}
    \centering
    \includegraphics[width=0.8\textwidth]{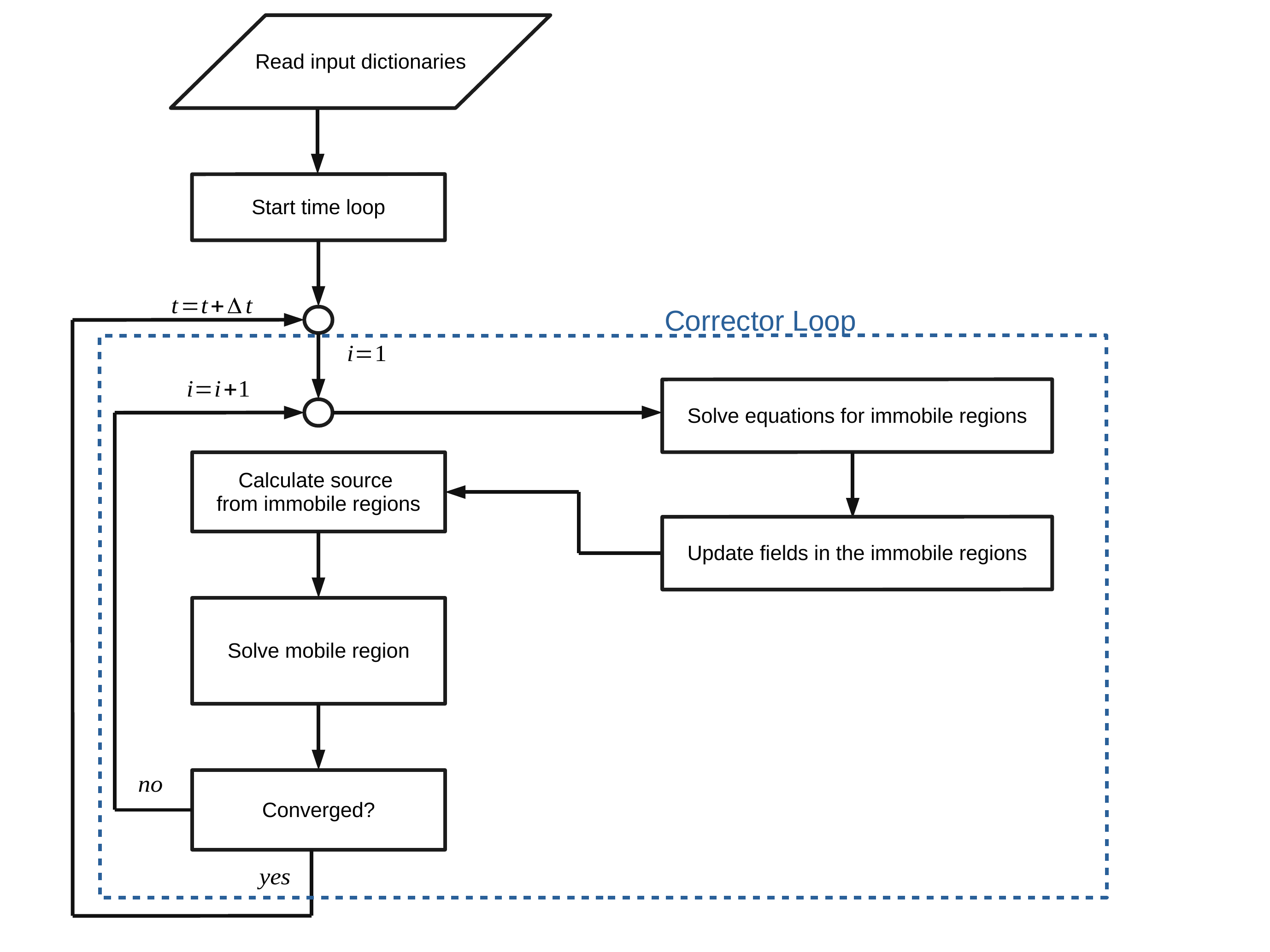}
    \caption{Diagram showing the numerical algorithm for \emph{multiRateScalarTransportFoam}.}
    \label{fig::mrmt_flow}
\end{figure*}

 The complete solution algorithm (including operations performed by the \emph{multiContinuumModel} library) is illustrated in Figure \ref{fig::mrmt_flow}, and it consists in a time loop with a nested corrector loop possessing a sub-time stepping loop. These operations can be summarised as follows:
 \begin{itemize}
     \item \textbf{Corrector loop}: this is the principal solution step. It consists in solving the governing equations for the immobile regions (with the optional sub-time stepping) and the advection-diffusion equation for the mobile region sequentially, in a segregated manner. When the residuals fall below a certain threshold or the maximum number of iterations (defined by the user) has been reached, the solver exits the loop.
     \item \textbf{Time loop}: it constitutes the main loop. After a satisfactory solution has been achieved in the corrector loop, the algorithm moves on to the next time step. 
 \end{itemize}
 
\subsection{Input files}
\label{S::input}

Input data and controls are provided by mean of appropriate 'dictionaries' (i.e., input files in \textsc{OpenFOAM}\textsuperscript{\textregistered} terminology). In the \emph{multiContinuumModel} library, input parameters must be provided in the \emph{multiRateProperties} dictionary located in the folder \emph{constant} (see the \textsc{OpenFOAM}\textsuperscript{\textregistered} \citep{Openfoam2019} details regarding the structure of simulation folders). 

\subsubsection{multiRateProperties dictionary}\label{sec:dictmultir}

Listing \ref{lst::mrp} shows an example of basic input for the \emph{multiRateProperties} dictionary. The first lines (1 to 8) are the required \textsc{OpenFOAM}\textsuperscript{\textregistered} header, and are present in all dictionaries in a similar fashion.
Lines 11 and 12 control the time-step adaptivity. In fact, the library allows to automatically set the time step $\Delta t$ in such way that the mass-transfer-based Courant number satisfies:

\begin{equation}
    \label{eq::reactCo}
    Co_{\lambda} = \text{max} \left( \alpha_{ik} \omega  \right) \Delta t < Co_{\text{max}}\,,
\end{equation}
 where $Co_{\text{max}}$ is the read from line 12. Ensuring that $Co_{\lambda} < 1$ is a necessary condition to obtain accurate and bounded results and therefore it is highly advised to keep this option active. However, in many applications (as groundwater transport) the time scale corresponding to the transport in the mobile region will be much smaller than the time scale of the mobile-immobile transfer. Therefore, in most circumstances the time step will not need to be adjusted.  
 
 Lines 14 to 24 of listing \ref{lst::mrp} contain the list of immobile regions. Each region is specified as a sub-dictionary with a user-defined name, which will be used by the library to identify the region and build the appropriate fields. Lines 16 to 23 show one immobile region as example. All immobile regions have the following entries:
 
 \begin{itemize}
     \item type: this is the type of immobile region, which defines how the $\alpha_{ik}$ and $\beta_{ik}$ are computed.
     \item numberOfTermsInExpansion: defines how many terms should be retained in the expansion.
     \item rescaleBetas: this entry allows to decide how the truncation of the series is handled. If set to true, all $\beta_{ik}$ are rescaled such that $\sum_k \beta_{ik} = 1$ as in \cite{Silva2009}. By default, this entry is set to false and a 'truncation capacity' $\beta_{tr} = 1 - \sum_k \beta_{ik} \neq 0$ will be computed. This truncation capacity will then be added to the capacity of the mobile region, as if the truncated terms were in equilibrium with the mobile concentration \cite{Municchi2020}.
 \end{itemize}

\begin{lstlisting}[caption={ Basic input for the \emph{multiRateProperties} dictionary}, captionpos=b, label={lst::mrp},style=customc,frame=single]
FoamFile
{
    version     2.0;
    format      ascii;
    class       dictionary;
    location    "constant";
    object      multiRateProperties;
}

//- Controls for time-step adaptivity
adaptiveTimeStepping    true;
maxCo                   0.95;

immobileRegions //- List of immobile regions
(
    nameOfRegion //- This can be any name
    {
        //- Type of immobile region
        type                    Spheres;
        
        //- Terms to represent the immobile region
        numberOfTermsInExpansion     50;
        
        //- How the truncation is handled
        rescaleBetas              false;
    }
);
\end{lstlisting}


\subsubsection{Immobile regions}

At the current stage, there are four different immobile region types available: Spheres, Cylinders, Layers, and a region for which mass transfer is given by a linear first-order process. We refer to the latter a first-order region.  
Table \ref{table::alphabeta} shows how the values of $\alpha_{ik}$ and $\beta_{ik}$  computed for different immobile regions. Since the zeros of the Bessel function are not computed explicitly, but are stored in an array, cylindrical immobile regions are limited to 50 terms in the expansion. First-order regions are the most flexible kind of immobile region, since the values of $\alpha_{ik}$ and $\beta_{ik}$ are read directly from the dictionary. This makes this type of immobile region appropriate for calibration studies. 

\begin{table}[]
\centering
\begin{tabular}{c c c } 
 \hline
  Type & $\alpha_{ik}$ & $\beta_{ik}$  \\  
 \hline\\
 Layers & \(\displaystyle \frac{\left(2k -1 \right)^2\pi^2}{4} \) & \(\displaystyle\frac{8}{\left(2k -1 \right)^2 \pi^2}\) \\&&\\
 Cylinders & \(\displaystyle \zeta_k^2 \) & $\displaystyle \frac{4}{\zeta_k^2}$ \\
 &&\\
 Spheres & $\displaystyle k^2\pi^2 $ & $ \displaystyle \frac{6}{k^2\pi^2}$ \\
 &&\\
 FirstOrderRegions & alphaCoeffs & betaCoeffs \\ [2ex]
 \hline
\end{tabular}
\caption{Expressions for $\alpha_{ik}$ and $\beta_{ik}$ for different types of immobile regions. Here $\zeta_k$ corresponds to the $k$-th zero of the 0-Bessel function of the first kind. The eigenvalues $\lambda_{ik}$ can be obtained using \cref{eq:dimless} and the $\omega$ reported in \cref{table:7Sp,table:composite}}
\label{table::alphabeta}
\end{table}

Listing \ref{lst::for} shows the syntax for a first-order region. Notice that the number of entries in alphaCoeffs and betaCoeffs must be equal to numberOfTermsInExpansion.

\begin{lstlisting}[caption={ Example of a FirstOrderRegions immobile region (named 'immobile' ) with three terms}, captionpos=b, label={lst::for},style=customc,frame=single]

    immobile
    {
        type                        FirstOrderRegions;
        numberOfTermsInExpansion                    3;
        alphaCoeffs                nonuniform (1 2 3);
        betaCoeffs           nonuniform (0.5 0.3 0.2);
    }

\end{lstlisting}

In addition to being defined in \emph{multiRateProperties}, immobile regions require a set of fields representing the initial condition, the capacity, and the transfer rate. These fields can be summarised (for each immobile region) as:

\begin{itemize}
    \item c.<name of region>: is the field representing the field $c$ (Favre averaged concentration) in that immobile region. It is possible to specify an initial condition for this field following the standard \textsc{OpenFOAM}\textsuperscript{\textregistered} syntax \cite{Openfoam2019}.
    \item omega.<name of region>: is the field representing the transfer rate for that immobile region. This field represents the material properties related to a immobile region and can vary in space and time. Specifically, it coincides with the ration between the diffusion coefficient and the square of a reference length for all region types except FirstOrderRegions.
    \item beta.<name of the region>: represents the capacity of the immobile region, and can vary in space and time.
\end{itemize}

The \emph{tutorials} folder in the library \citep{federico_municchi_2020_3938868} provides a range of examples illustrating the syntax and how to structure a simulation folder.

\section{Results and discussion}


In this Section we show applications of our library in different situations taken from problem settings in heterogeneous porous media and flow in packed bed equipment. We choose the different systems mainly to show the the flexibility of our model and implementation in different situations, but they also have the purpose to show the wide range of applicability of such a model. We start from a simple $0D$ model and then we move to more realistic-like cases in following sections.

One important numerical parameter investigated is $M$, the number of terms of expansions terms, (i.e. we truncate the infinite series of eigenfunctions in \cref{eq::mrmt_disc} after $M$ terms). The truncation is handled differently depending on the flag \texttt{rescaleBetas}, as described in \cref{sec:dictmultir}. 
In \cref{table:summary} we summarise all the cases we consider in the following specifying the different geometries considered along with their relevant initial and boundary conditions. These are particularly important as they define the kind of mass-transfer happening in the system and the effect of the immobile regions. Most cases are solved for empty immobile regions and fully saturated mobile regions, resulting in a maximum transfer in the initial transient. In presence of an inflow with concentration zero (1D,  2D, 3D), the initial mass in the system is all flushed away with the immobile regions slowing down the process, storing temporarily some mass. The stationary state in this case, is when the system is completely empty with all concentrations equal to zero.
All the cases we present in this section are included in the library package as tutorials.

\begin{table}[]
\centering
\begin{tabular}{c c c c} 
\specialrule{.2em}{.1em}{.1em} 
\multicolumn{4}{c}{0D}\\
   \hline
  &  $c_{m}(t=0)$ & $c_{im}(t=0)$ & $c_{m,inlet}$ \\  
 \hline
\textsc{7Sp} & $1$ & $0$ & - \\
\specialrule{.2em}{.1em}{.1em} 
\multicolumn{4}{c}{1D}\\
   \hline
  &  $c_{m}(t=0)$ & $c_{im}(t=0)$& $c_{m,inlet}$ \\  
     \hline
First-order & 0 & 1 & 0\\
\textsc{1Sp}  & $1$ & $0$  & $0$\\ 
\specialrule{.2em}{.1em}{.1em} 
\multicolumn{4}{c}{2D}\\
   \hline
  &  $c_{m}(t=0)$ & $c_{im}(t=0)$& $c_{m,inlet}$ \\  
     \hline
\textsc{7Sp} & $1$ & $0$  & $0$\\ 
\textsc{Comp}  & $1$ & $0$  & $0$\\ 
\textsc{Rand2D}  & $1$ & $0$  & $0$\\ 
\specialrule{.2em}{.1em}{.1em} 
\multicolumn{4}{c}{3D}\\
   \hline
  &  $c_{m}(t=0)$ & $c_{im}(t=0)$& $c_{m,inlet}$ \\  
     \hline
\textsc{7Sp}  & $1$ & $0$  & $0$\\ 
\specialrule{.2em}{.1em}{.1em} 
\end{tabular}
\caption{Summary of the setup of the cases considered here. $c_{m}(t=0)$ and $c_{im}(t=0)$  are the initial concentrations of the mobile and immobile regions respectively, inside the domain.  $c_{m,inlet}$ is the boundary condition at the inlet for the mobile concentration. }
\label{table:summary}
\end{table}

\subsection{Zero-dimensional test-case}\label{sec:7sphere}

For the zero-dimensional test-case, we solve \cref{eq::mrmt_fvm} in \OF in a domain composed by a single cell. In this way, we obtain a configuration akin to that described in \citfull{Haggerty1995} for batch reactors (no advection, dispersion, sources and sinks). We choose to represent the immobile region as composed by seven spherical immobile regions, and we called this the \textsc{7Sp} model. This is analogous to the one reported in Tab.2 of \citfull{Haggerty1995}, with the same $\beta$ and $\omega$ which we report in \cref{table:7Sp}. Our purpose here is to compare our results with those of \citfull{Haggerty1995} to demonstrate the accuracy of our implementation. 

\begin{table}[]
\centering
\begin{tabular}{c c c } 
 \hline
  &  $\omega_i$ (s$^{-1}$) & $\beta_i$  \\  
 \hline\\
Sphere1 & $3.1\cdot10^{-8}$& $0.0406$\\
Sphere2 & $9.2\cdot10^{-8}$& $0.1699$\\
Sphere3 & $2.3\cdot10^{-7}$& $0.2731$\\
Sphere4 & $2.7\cdot10^{-7}$& $0.2592$\\
Sphere5 & $9.4\cdot10^{-7}$& $0.1548$\\
Sphere6 & $1.7\cdot10^{-6}$& $0.0620$\\
Sphere7 & $1.4\cdot10^{-6}$& $0.0404$\\
 \hline
\end{tabular}
\caption{Values of $\omega_i$ (where $i=1,\ldots,7$) and capacity coefficients $\beta_i$ as reported in Tab.2 of \citfull{Haggerty1995} used for the calculation of the \textsc{7Sp} model. }
\label{table:7Sp}
\end{table}

\begin{figure}
    \centering
    \includegraphics[width=0.6\textwidth]{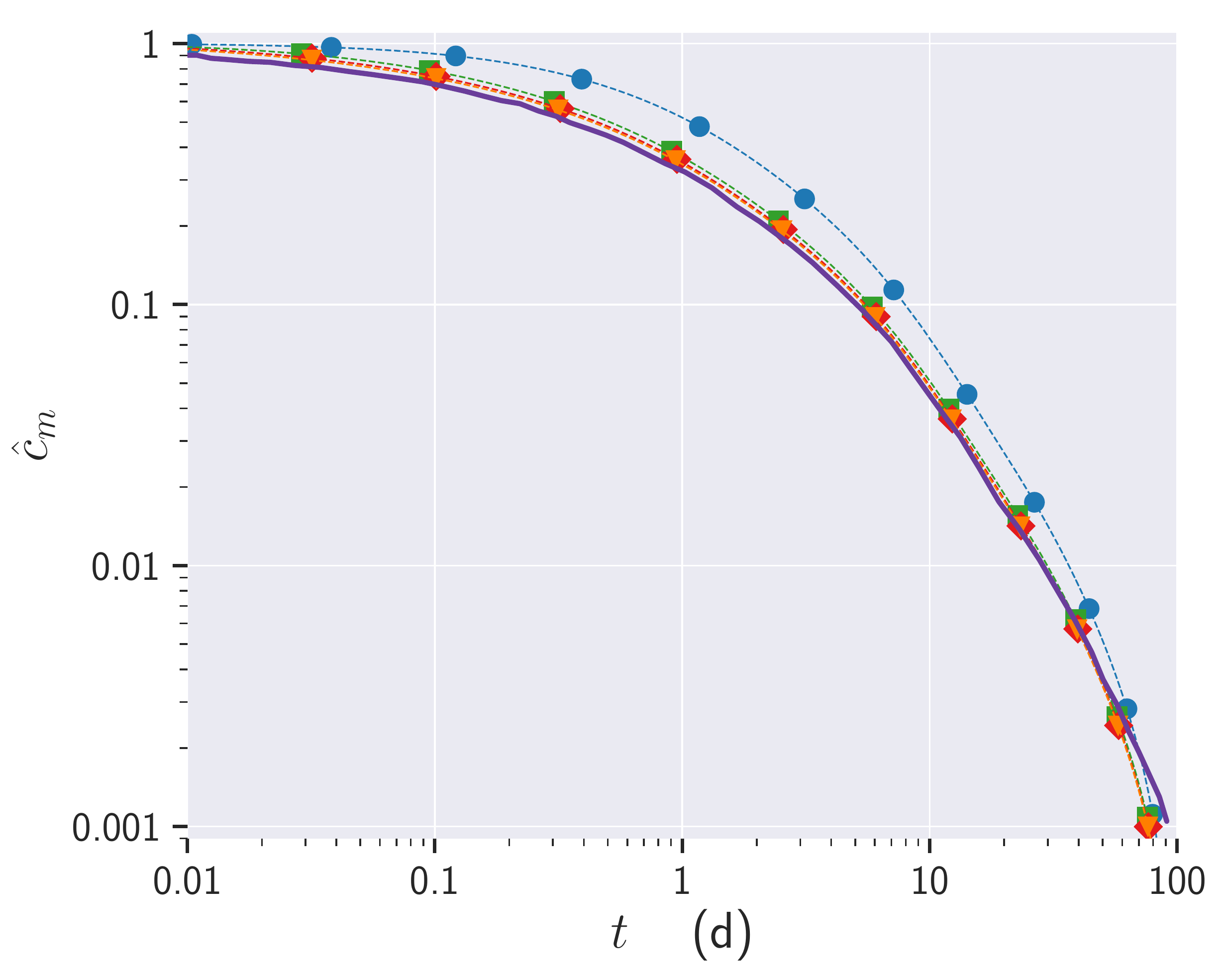}
    \caption{Breakthrough curves at different expansion for the \textsc{7Sp} model (see \cref{table:7Sp})  compared with the same model reported in \citfull{Haggerty1995}. The purple continuous line is taken from \citep{Haggerty1995} (Table 2), blue $\bullet$ $M$=2, green $\blacksquare$ $M=10$, red $\blacklozenge$ $M=20$, orange $\blacktriangledown$ $M=50$  
    }
    \label{figure:7Sp}
\end{figure}

We report the results for four different numbers of retained eigenfunctions in the expansion, identified by $M=2,10,20,50$, in terms of the normalised concentration of the mobile region $\hat{c}_m$, defined as
\begin{equation}\label{ctilde}
\hat{c}_m=\frac{c_{eq}-c_m(t)}{c_{eq}-c_{m}(t=0)}\,,
\end{equation}
where $c_{eq}$ is  the equilibrium concentration between mobile and immobile regions, $c_{m}(t=0)$ is the initial concentration in the mobile regions, $c_m(t)$ concentration in the mobile region at time $t$.
The curve we obtain is shown in \cref{figure:7Sp}. As it can be seen, using only two terms (i.e. $M=2$) is not sufficient to capture the dynamics of the process. However, $M=10$ seems already enough to obtain good quantitative agreement with the known results (see Figure 2-b in \citep{Haggerty1995}). As expected, the agreement increases as we increase the number of expansions.
However, notice that our method is prescribing a slightly different trend due to the way we account for the truncated terms (without rescaling), which is not detailed in \citep{Haggerty1995}.

\clearpage
\subsection{One-dimensional test cases}


In the following, we discuss two one-dimensional test cases characterised by one single first-order and sphere regions.

\subsubsection{First-order immobile regions}
\label{ss:chebfun}
In order to assess the accuracy of our numerical method and the correct implementation of the library, we compare results from the \texttt{multiRateScalarTransportFoam} numerical solver described in \cref{S::multiRateScalarTransportFoam} against a spectral solution for a simple problem with advection and diffusion.
We therefore consider the following system of equations:
\begin{equation}
    \label{eq::chebProblem}
    \begin{dcases}
    \ddt{c_m} + \frac{\partial c_m}{\partial x} - \frac{1}{10}\frac{\partial^2 c_m}{\partial x^2} = \frac{1}{2}\left( c_i -c_m \right) \\
    \ddt{c_i} = \left( c_m -c_i \right)\,,
    \end{dcases}
\end{equation}
together with the following boundary conditions:
\begin{equation}
    \left.\frac{\partial c_i}{\partial x}\right|_{x=0} = \left.\frac{\partial c_i}{\partial x}\right|_{x=1}=0, \quad c_m(t,0) = 0, \quad \left.\frac{\partial c_m}{\partial x}\right|_{x=1}=0\,, 
\end{equation}
and initial conditions:
\begin{equation}
    c_m(x,t=0)=0, \quad c_i(x,t=0)=1\,.
\end{equation}

Notice that the coefficients appearing in system \ref{eq::chebProblem} are chosen arbitrarily to generate fast transients dominated by convection and mass transfer, and do not necessarily represent any realistic application to groundwater flows. However, they provide an excellent test for the numerical stability of this algorithm and the relatively large diffusion allows for efficient spectral solutions.

In this simplified mathematical model, at time $t=0$  mobile and immobile regions begin exchanging mass, starting from a condition of non-equilibrium. Specifically, mass is transferred from the immobile region to the mobile region and then transported out of the domain by advection. System \ref{eq::chebProblem} is solved using the Matlab-based library \textsc{Chebfun} \cite{Driscoll2014}, which employs Chebyshev polynomials to solve systems of differential equations to spectral accuracy, providing our benchmark solution. 
System \ref{eq::chebProblem} is discretised in \OF using a second order  schemes in space  (i.e., \texttt{linearUpwind} scheme for advection and \texttt{Gauss linear}, i.e., central differences, for diffusion) and in time (\texttt{backward} scheme). A grid of $20$ cells was used and the time step was chosen dynamically to satisfy the Courant-Friedrichs-Lewys condition $Co = U\Delta t / \Delta x < 1$ \cite{Courant1928}, where $Co$ is the Courant number, $U=1$, $\Delta t$ is the time step, and $\Delta x$ is the mesh spacing. Such time step was small enough to satisfy also the condition for $Co_{\lambda}$.

In the following, the quantity of interest is the normalised flux of concentration across the downstream  boundary of the domain, also called the breakthrough curve:
\begin{equation}
    \label{eq::BT}
    BT(t) = \frac{\displaystyle\oint_{S_{\text{out}}} c_m (\bm{x},t) \bm{u} \cdot \text{d}\bm{S}}{\displaystyle\oint_{S_{\text{out}}} 
    \bm{u} \cdot \text{d}\bm{S}} \,,
\end{equation}
where $S_{\text{out}}$ is the outlet boundary of the domain.
Furthermore, we will often look at the average concentration in the immobile regions within the domain defined as:

\begin{equation}
    \label{eq::cimm_def}
    c_{im} (t) = \frac{\displaystyle \int_{\Omega}c_i \ofxt \text{d}V}{\displaystyle \int_{\Omega} \text{d}V}\,.
\end{equation}

Notice that $c_{im}$ can be computed for different immobile regions. However, we will not assign a subscript $i$ to $c_{im}$. Instead, we will specify in each plot which is the immobile regions we are considering.  

\begin{figure}
    \centering
    \includegraphics[width=0.4\textwidth]{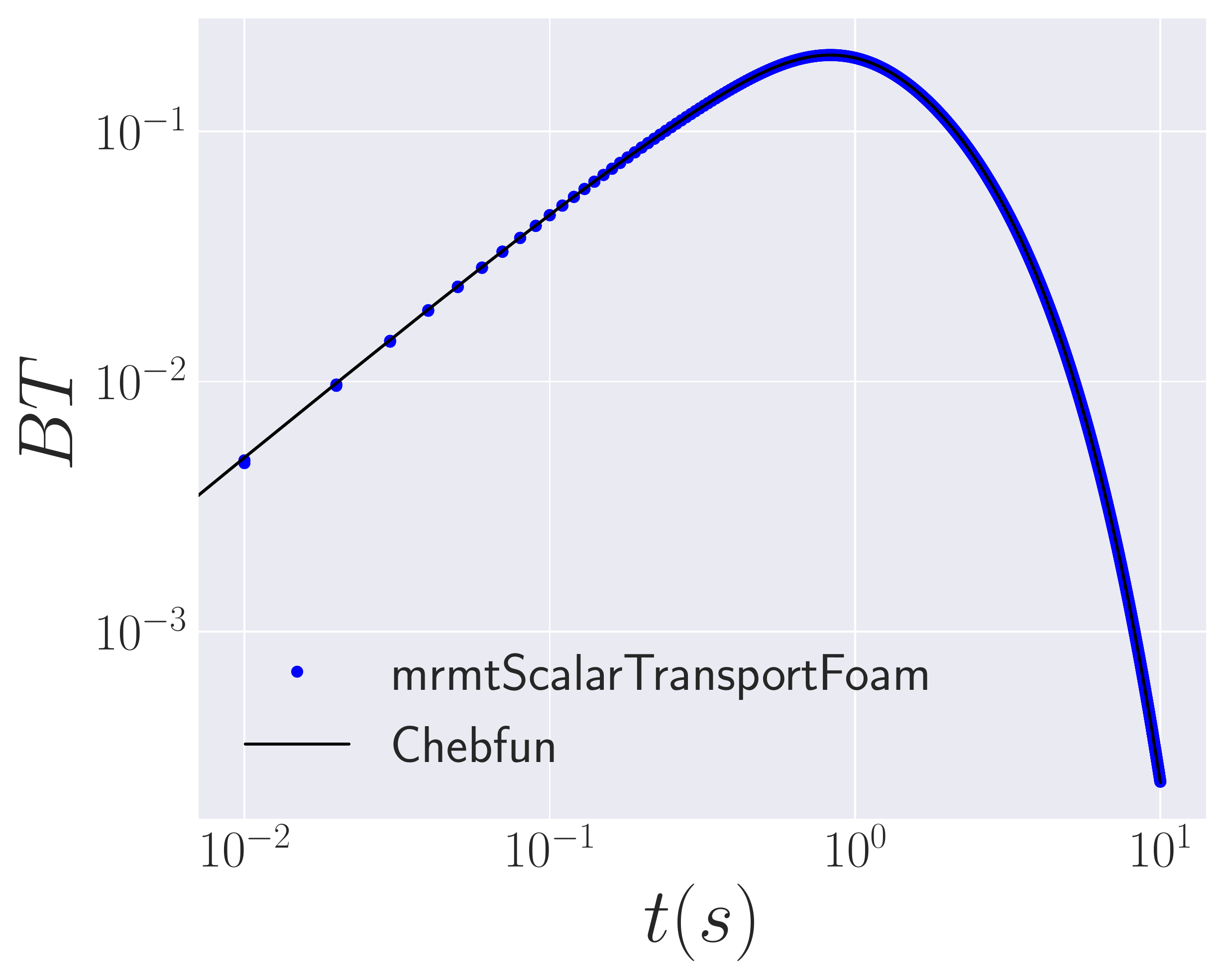}
    \caption{Breakthrough curves from \textsc{Chebfun} \cite{Driscoll2014} and our \textsc{OpenFOAM}\reg solver \texttt{multiRateScalarTransportFoam}. The root mean square deviation between the two methods is $0.33\%$.} 
    \label{fig:verification}
\end{figure}

As it can be seen in Figure \ref{fig:verification}, there is an excellent agreement between the breakthrough curve predicted by \textsc{Chebfun} and that predicted by \texttt{multiRateScalarTransportFoam}. 
In the initial times, there is a sharp increase in the breakthrough curve cause by  concentration being mobilised and flushed out. After reaching a peak, the diffusion out of the immobile region and subsequent advection-diffusion out of the domain decays exponentially to zero. This is expected as we used a single-rate mass transfer model.


\subsubsection{Spherical immobile regions}

In this testcase, we solve again system  \ref{eq::chebProblem}, replacing the RHS single-rate mass transfer with the multi-rate expansion of a single sphere. Therefore we denote it by \textsc{1Sp}.
Contrarily  to the previous case, the immobile regions  start here completely empty, and all the concentration is initially in the mobile part of the system.

 \begin{figure*}
  \begin{subfigure}[b]{0.5\textwidth}
   \centering
    \includegraphics[width=0.8\textwidth]{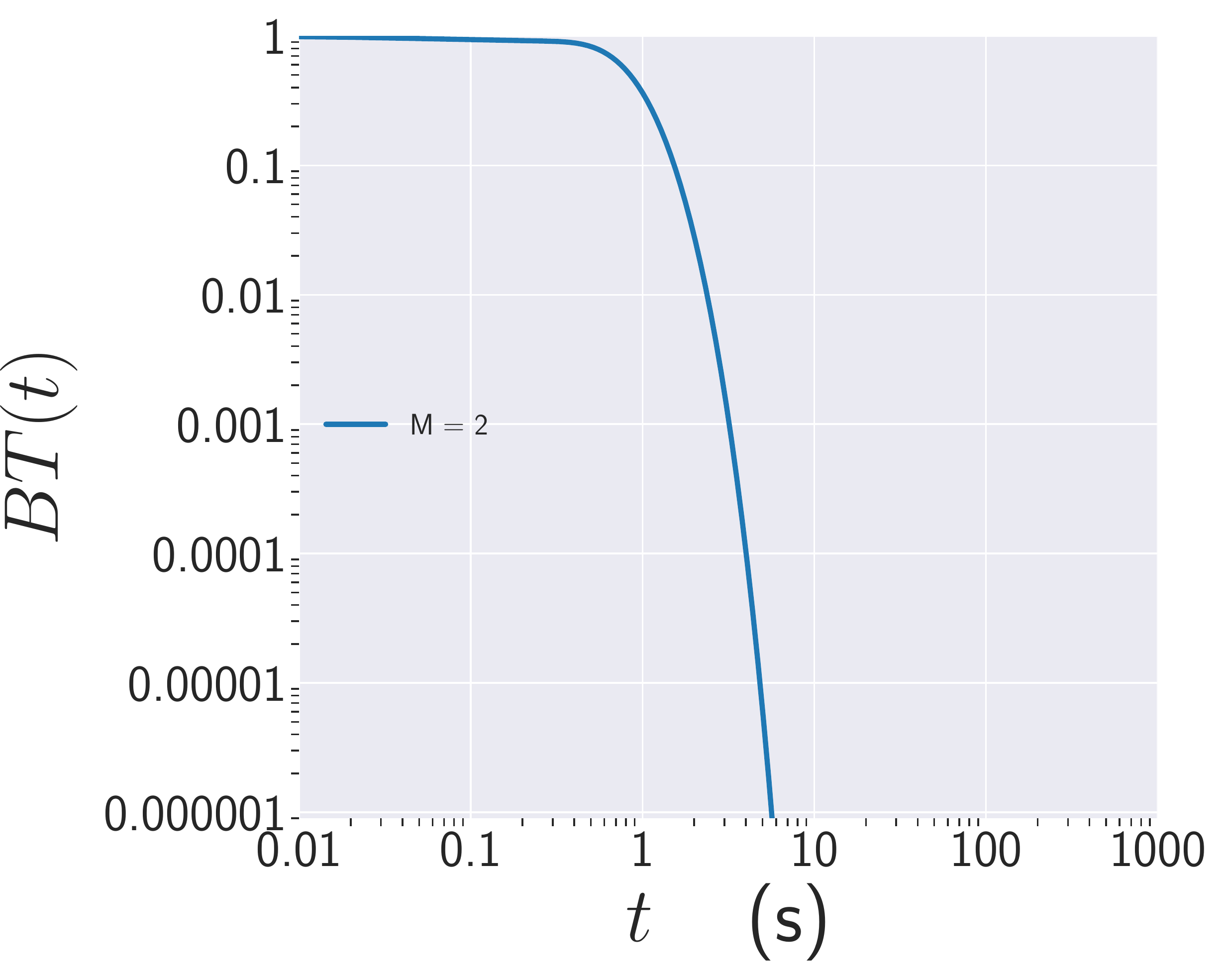}
    \caption{ $\omega=1.0 \;  \text{s}^{-1} $, $\beta=0.1\, \beta_m$} 
    \label{fig:rates_a} 
    \vspace{2ex}
  \end{subfigure}
 \begin{subfigure}[b]{0.5\textwidth}
    \centering
   \includegraphics[width=0.8\textwidth]{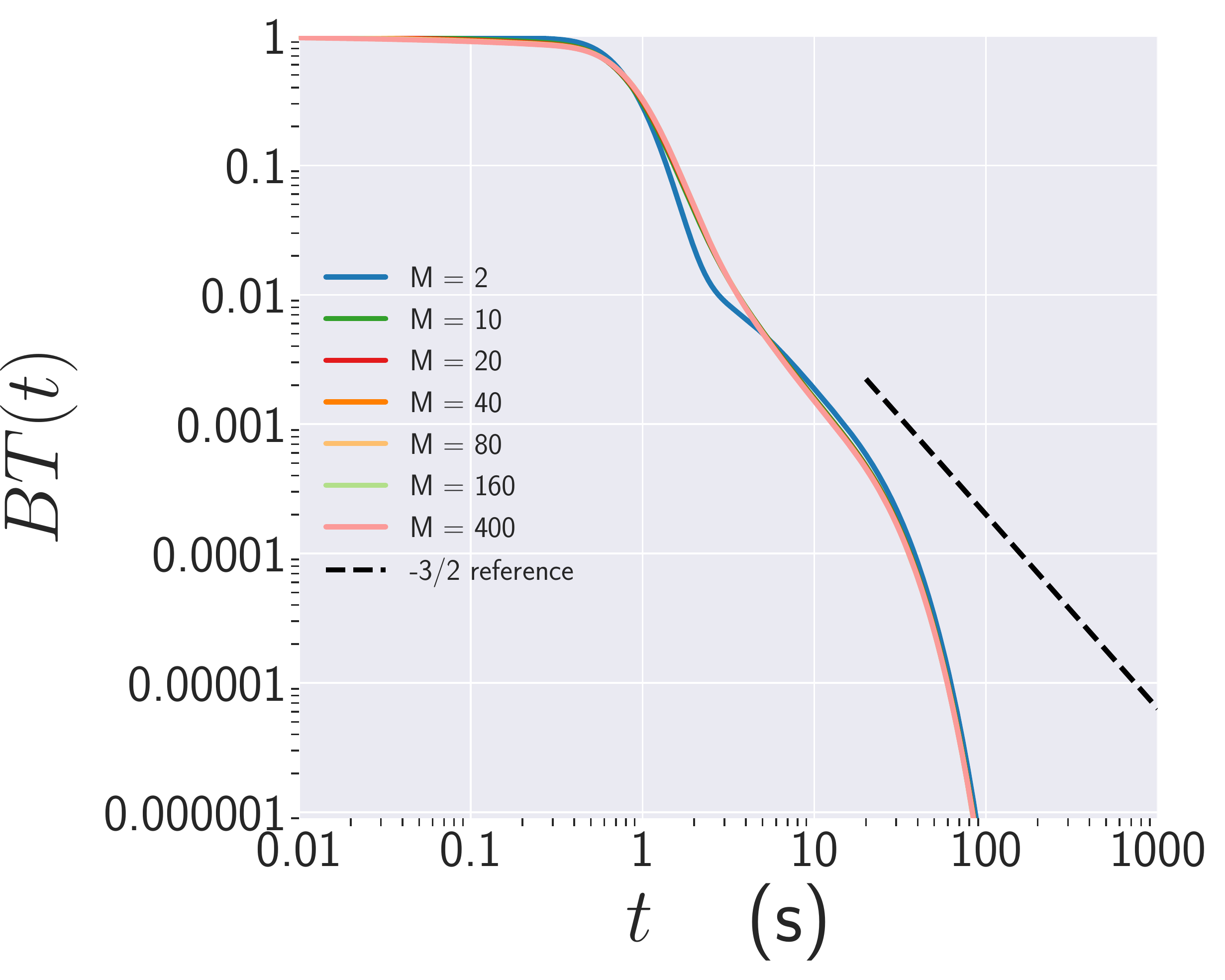}
    \caption{ $\omega=0.01 \;  \text{s}^{-1} $, $\beta=1 \beta_m$} 
     \label{fig:rates_b} 
    \vspace{2ex}
  \end{subfigure} 
    \begin{subfigure}[b]{0.5\textwidth}
   \centering
    \includegraphics[width=0.8\textwidth]{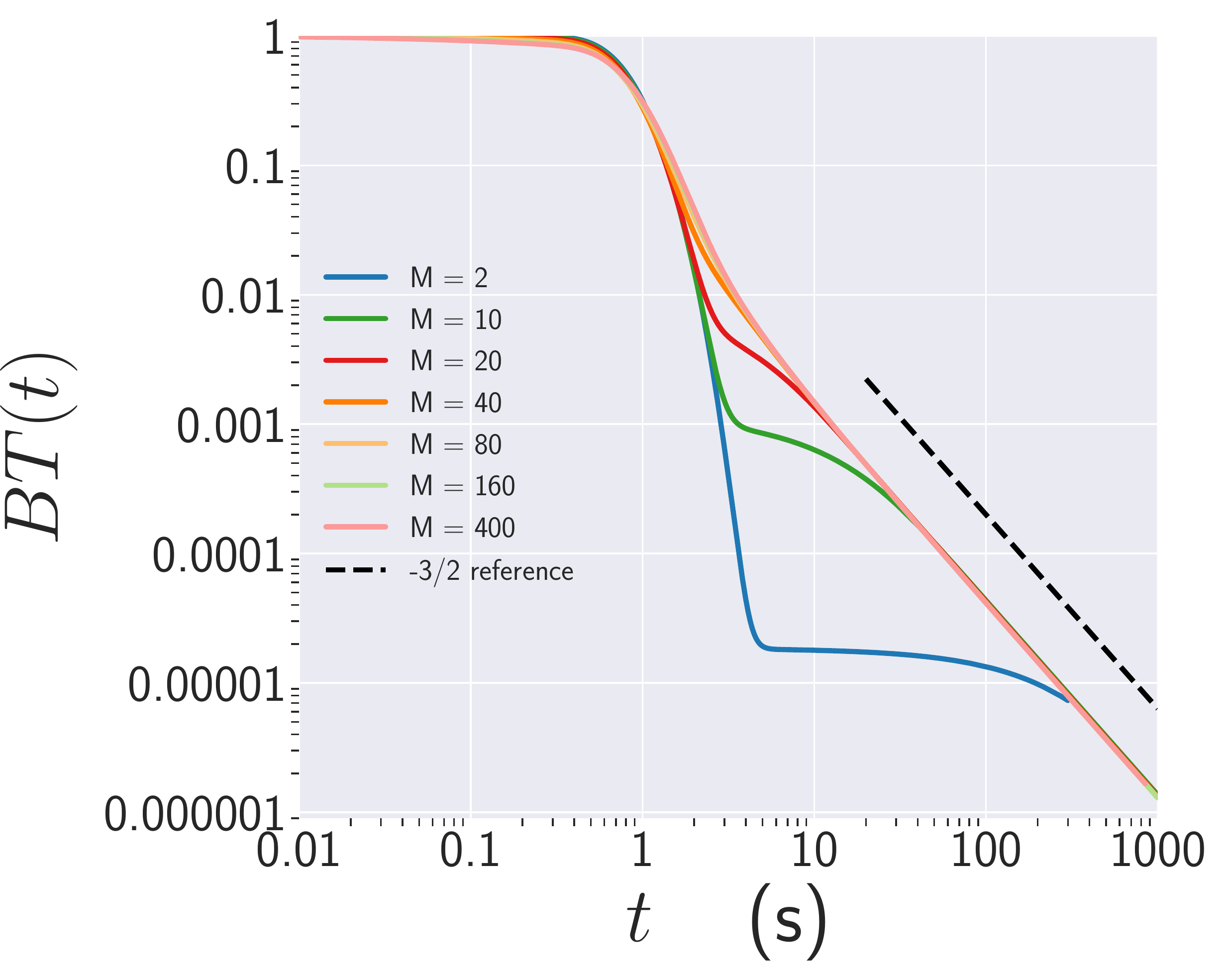}
    \caption{ $\omega=10^{-4}\;  \text{s}^{-1} $, $\beta=10\, \beta_m$} 
    \label{fig:rates_c} 
    \vspace{2ex}
  \end{subfigure}
 \begin{subfigure}[b]{0.5\textwidth}
    \centering
   \includegraphics[width=0.8\textwidth]{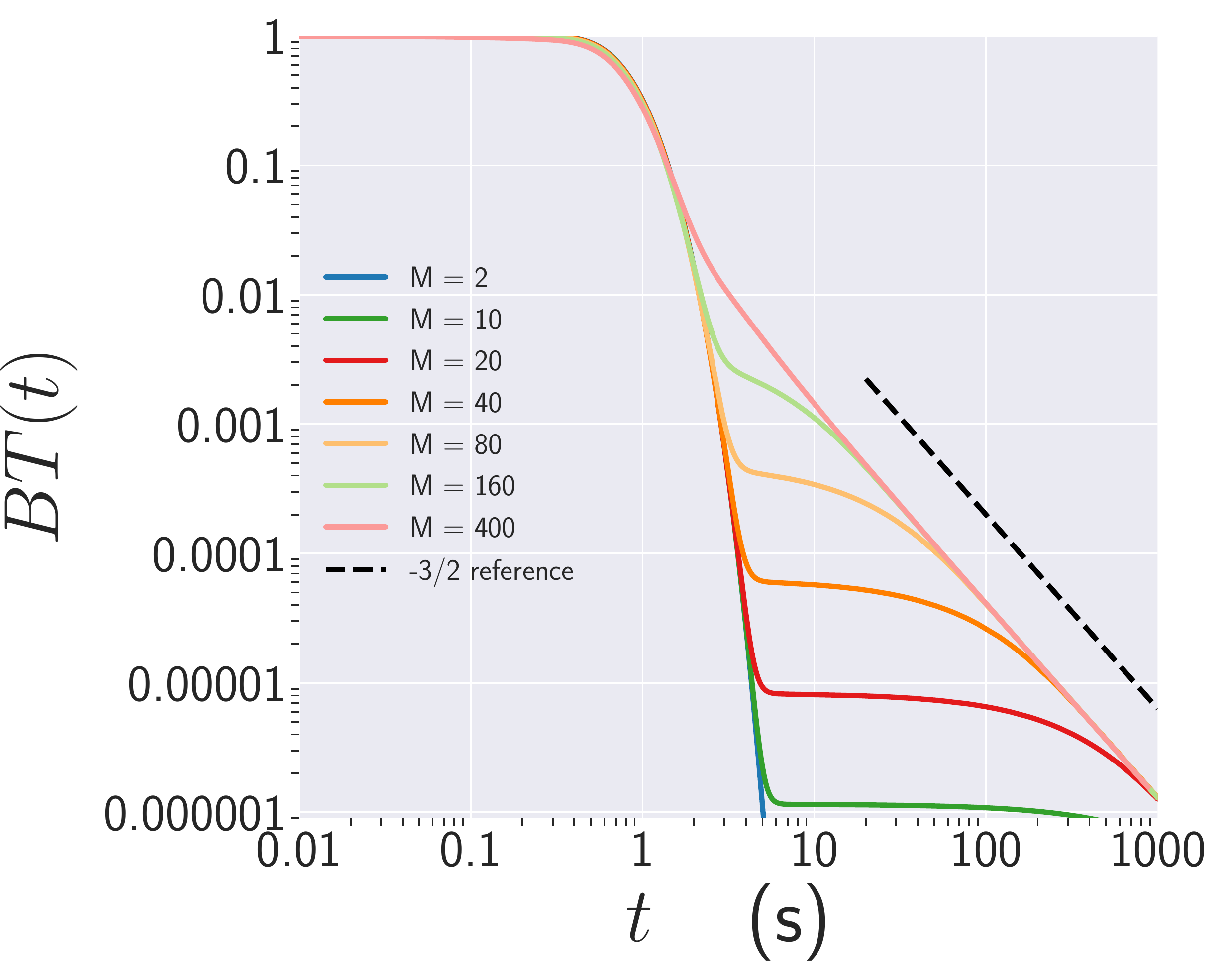}
    \caption{ $\omega=10^{-6}\;  \text{s}^{-1} $, $\beta=100\, \beta_m$} 
     \label{fig:rates_d} 
    \vspace{2ex}
  \end{subfigure} 
 \caption{  Breakthrough curves for the \textsc{1Sp} model at different values of $\omega$ and $\beta$ for different expansion numbers. The top-left plot reports just a single curve since they are all coincident}
    \label{fig:1Dcurves}
\end{figure*}

Results are presented in figure \ref{fig:1Dcurves} for various numbers of terms in the eigenfunction expansions, and different values of $omega$ and $beta$ for the spherical inclusions. Other simulations parameters are identical to those in \ref{ss:chebfun}. 

Figure \ref{fig:rates_a} shows the results for $\omega=1.0$ and $\beta=0.1$, which represent the maximum value of $\omega$ and the minimum value of $\beta$ we probed. Such values result in very large eigenvalues corresponding to characteristic response times $\tau_{k} = \alpha_{k} \omega^{-1} > 1$, which leads to  a very fast mass transfer that is dominated by  leading order term in the expansion. Furthermore, the low value of $\beta$ (compared to the size of the mobile region) means that the spheres have low capacity. Therefore, it is expected that a small number of terms in the expansion would be enough to capture the system dynamics, which consists in a fast evolution with little or negligible non-local effects. This is precisely what is observed in figure \ref{fig:rates_a} (where increasing the number of terms does not result in any significant change and are therefore omitted). 

Conversely, decreasing $\omega$ leads to slower transients due to the slower response time of the modes. Therefore, the role of time history on the system dynamics increases from figure \ref{fig:rates_b} to \ref{fig:rates_d}, and a larger number of terms is required to capture the fast transients. In fact, the leading order term corresponds to the slowest dynamics (the smallest eigenvalue) and it is often the only one retained in asymptotic theories (where the initial transient is disregarded). It is however necessary to retain a large number of terms in order to accurately predict non-equilibrium dynamics. It should be noted that the capacity is also playing an important role.  When the capacity of the system increases and approaches the limit of infinite capacity (i.e., infinite size) of the immobile regions the slope of the breakthrough approaches $-3/2$ \citep{Kekalainen2011} as shown in figure \ref{fig:rates_d}.

\clearpage
\subsection{Two-dimensional simulation of heterogeneous porous media} \label{Sec:2D}

In this section, we apply \texttt{multiRateScalarTransportFoam} to solve for solute transport in media characterised by spatially variable properties. The modelling of transport of dissolved substances and energy in heterogeneous porous media is a key issue in a series of applications ranging from groundwater remediation~\cite{Domenico1997} to radionuclide migration~\cite{Geckeis2012}, and the geological storage of carbon dioxide~\cite{Niemi2017}. 

The computational domain considered here is a 2D domain of width 2 m and height 1 m. The numerical grid was built in \OF  using the \texttt{blockmesh} utility, which allows the generation of orthogonal hexahedral meshes. The total number of cell in our computational domain is 20000.

\subsubsection{Flow and permeability fields}

\begin{figure*}
  \begin{subfigure}[b]{0.5\textwidth}
   \centering
    \includegraphics[width=1.01\textwidth]{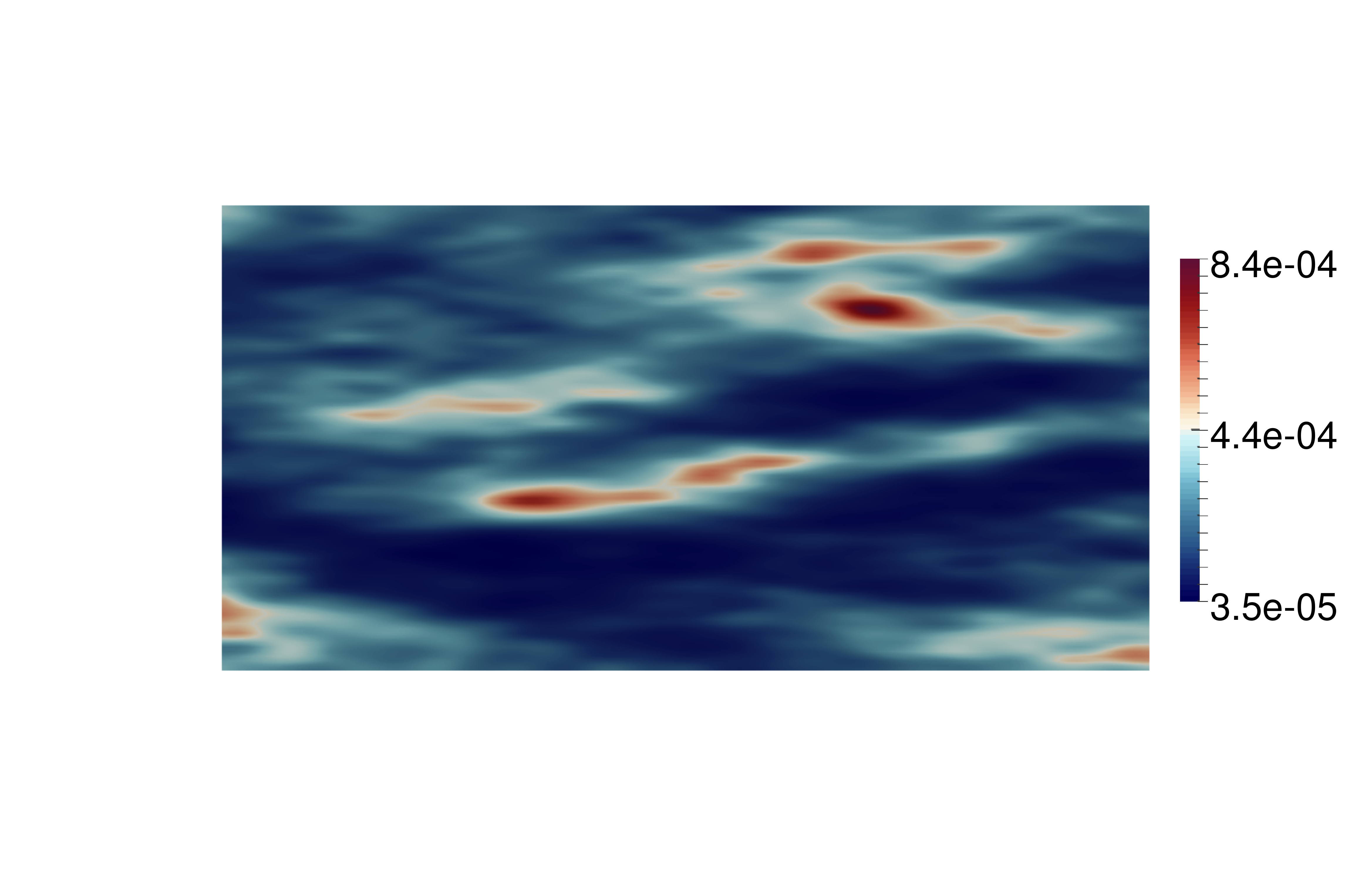}
        \vspace{-6ex}
    \caption{ Permeability Field (m$^2$)} 
    \label{fig:Krand} 
    \vspace{2ex}
  \end{subfigure}
 \begin{subfigure}[b]{0.5\textwidth}
    \centering
   \includegraphics[width=0.8\textwidth]{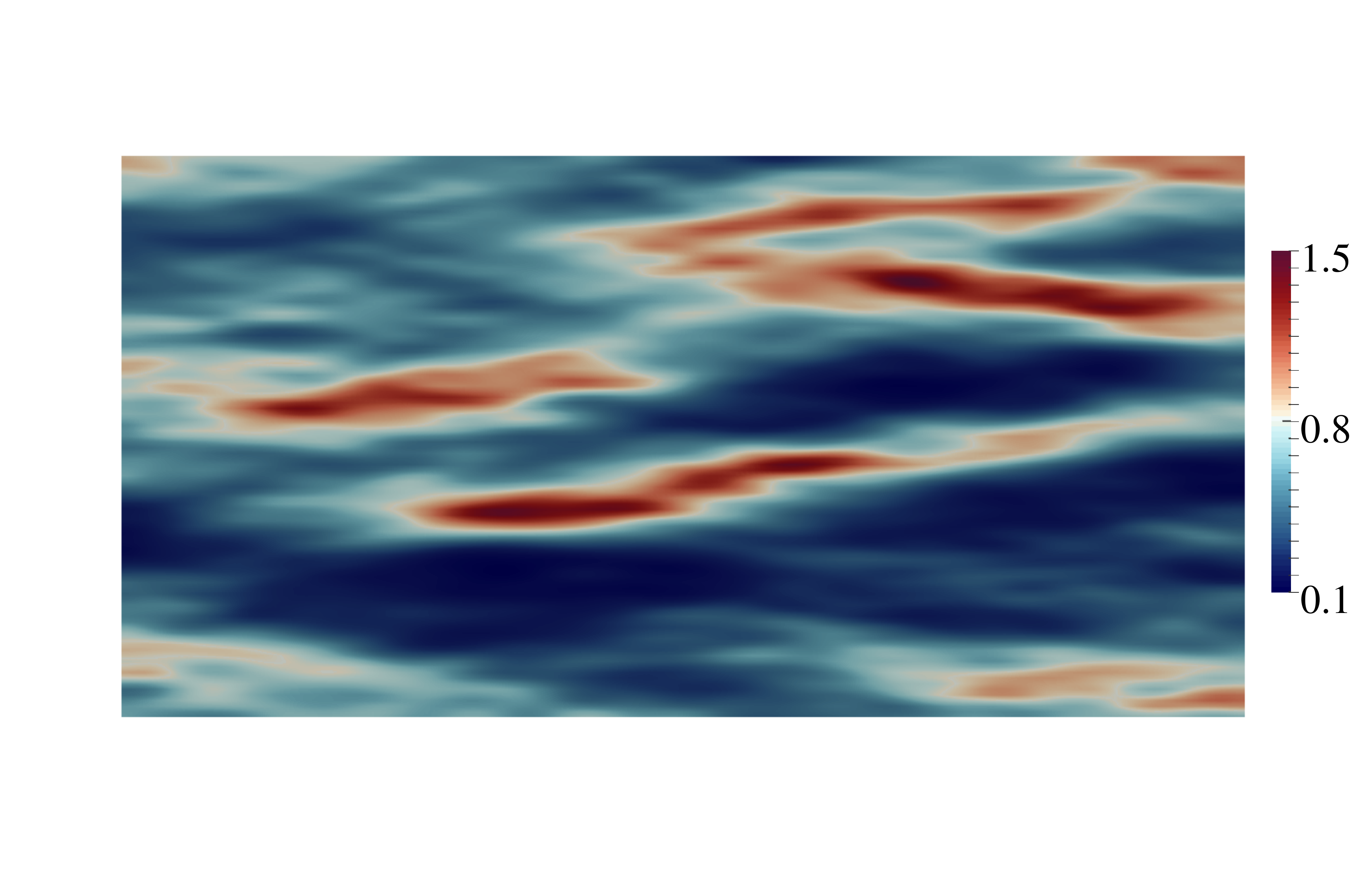}
    \caption{ Velocity Field (m/d)} 
     \label{fig:UR} 
    \vspace{2ex}
  \end{subfigure} 
 \caption{  Contour plot of the randomly generated permeability field and the corresponding velocity field obtained by solving \cref{Eq:Darcy}}
    \label{fig:U-Krand}
\end{figure*}

In all the subsequent cases, the flow field in the porous media is obtained by solving the steady-state Darcy equation:
\begin{equation}\label{Eq:Darcy}
    - \mu^{-1} \mathbf{K} \cdot \grad P  = \bm{u} \, , \quad
    \grad \cdot \bm{u}  = 0 \,,
\end{equation}
where $\mu$ is the dynamic viscosity of the fluid, $K$ is the permeability tensor, and $P$ is the pressure. In this work we considered the dynamic viscosity of water at 298 K, $\mu = 8.9\cdot 10^{-4}\,\,\,$ Pa s. Equation \ref{Eq:Darcy} is solved with boundary conditions on the pressure imposing a pressure drop $\Delta P$ between the two ends of the computational domain. 
From \cref{Eq:Darcy} we obtain a Poisson equation whose solution is easily implemented in \OF.
The velocity is the computed from the fluxes of the Poisson equation. This last passage is the finite volume equivalent of calculating $\bm{u}$ using equation \ref{Eq:Darcy} (with the solenoidal velocity condition). 
In general, the permeability field $\mathbf{K}$ is an anisotropic tensorial field which can be constant in the domain, or spatially variable. 
We consider here an isotropic permeability field of the form $\mathbf{K}=k\ofx \mathbf{I}$, with $\mathbf{I}$ being the identity tensor and $k\ofx$ a Gaussian random field.
A key feature of natural and engineered porous media is, in fact, spatial heterogeneity. We model the spatially varying permeability as a realisation of a Log-Normal spatial random field, with $\log(k)$ a Gaussian random field with zero mean and unit variance. We assume an exponential correlation between the points with correlation length equal to 0.5 and 0.1, in the longitudinal and vertical  direction respectively. The resulting random field is shown in figure \ref{fig:Krand}.

%
%
%

We solve \cref{Eq:Darcy} with a pressure drop of $\Delta P=10^{-6} \; \text{Pa}$ (which translates to Dirichlet boundary conditions of $P(0,z)= 10^{-6} \; \text{Pa}$ and $P(L,z)=0 \; \text{Pa}$). The resulting flow-field for the permeability field shown in \ref{fig:Krand} in reported in \cref{fig:UR}. As expected, the higher velocities corresponds to regions with higher permeability $K$.

%
%
%
%

In the following we consider solute transport in theses spatially variable flow fields combined with mobile-immobile mass transfer characterised by constant and spatially variable properties. From a phenomenological point of view, mobile-immobile mass transfer can be considered to account for the impact of small-scale medium heterogeneities, while large scale variability is accounted for explicitly~\cite{cortis2004}.

For the  case of constant MRMT parameters, we will analyse  two different models: $i$) the \textsc{7Sp} model that we discussed in \cref{sec:7sphere} (see also \cref{table:7Sp}), $ii$) the \textit{Composite} (\textsc{Comp}) model summarised in \cref{table:composite}  that is a combination of  different immobile regions. For the  case of heterogeneous MRMT parameters,  the \textit{Random} (\textsc{Rand2D}) model consists  in one immobile spherical region for which all the relevant parameters $K$, $\beta_i$, and $\omega_{i}$ are non-uniform and dependent on $k$. 

\subsubsection{Homogeneous mass transfer properties}


We will mainly focus our discussion on the \textsc{7Sp} model and at the end of the section we will show some results for the \textsc{Comp} model.

\begin{table}[]
\centering
\begin{tabular}{c c c } 
 \hline
  &  $\omega_i$ (s$^{-1}$) & $\beta_i$  \\  
 \hline\\
Sphere1 & $2.80\cdot10^{-7}$& $0.35$\\
Sphere2 & $1.75\cdot10^{-8}$& $0.20$\\
Layer1 & $1.43\cdot10^{-9}$& $0.15$\\
Cylinder1 & $1.00\cdot10^{-9}$& $0.15$\\
FirstOrder1 & $2.76\cdot10^{-6}$& $0.05$\\
FirstOrder2 & $4.42\cdot10^{-5}$& $0.10$\\
 \hline
\end{tabular}
\caption{Values of $\omega_i$ (where $i=1,\ldots,7$) and capacity coefficients $\beta_i$ as reported in Tab.3 of \citfull{Haggerty1995} used for the calculation of the \textsc{Comp} model. }
\label{table:composite}
\end{table}

\noindent \textbf{{\textsc{7Sp}} Model:} 

\begin{figure}
  \begin{subfigure}[b]{0.5\textwidth}
   \centering
    \includegraphics[width=0.95\textwidth]{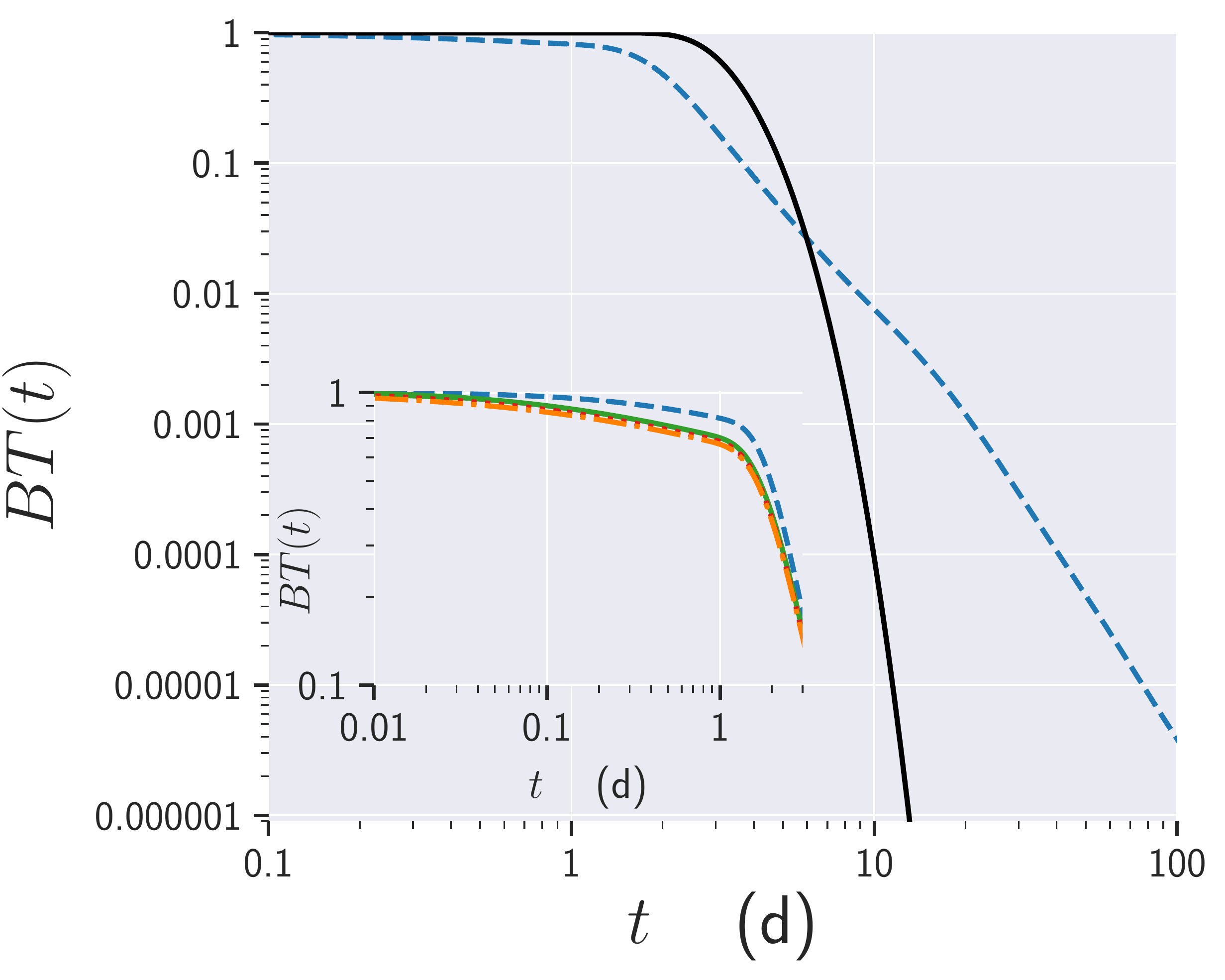}
    \caption{}
        \label{fig:BT2D}
  \end{subfigure}
 \begin{subfigure}[b]{0.5\textwidth}
    \centering
   \includegraphics[width=0.95\textwidth]{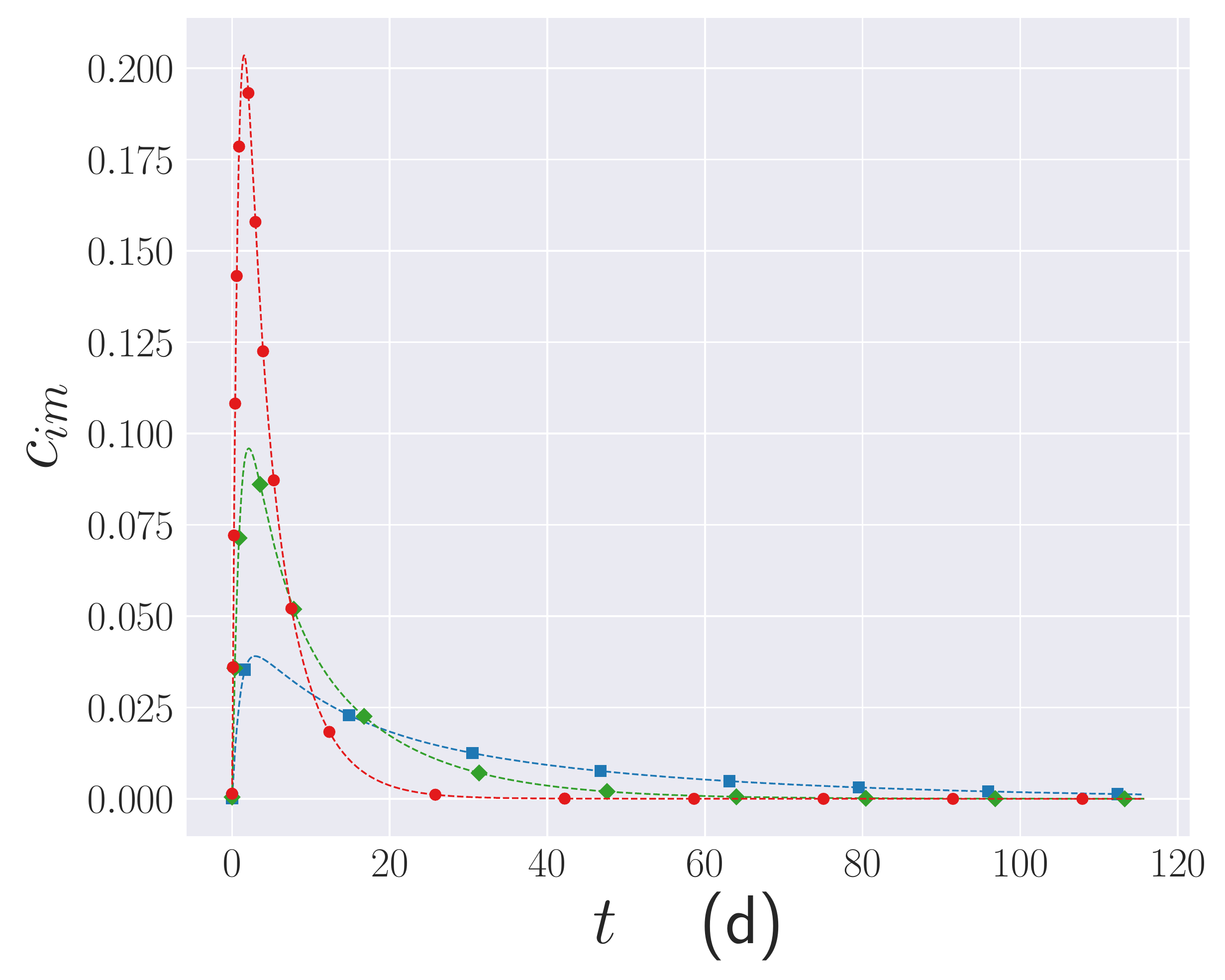}
    \caption{}
        \label{fig:7SpConc}
    \vspace{2ex}
  \end{subfigure} 
 \caption{On the left: breakthrough curves for the \textsc{7Sp} case and $M=2$ and comparison with the case without multi-rate (black continuous curve). In the inset, are reported the results for the first 3 days for all the expansion considered: $M=2$ dashed blue curve, $M=10$ green continuous curve, $M=20$ dotted curve, $M=50$ dash-dotted curve. On the right: concentration of the concentration versus time in three of the immobile regions in the Sp7 model.  Sphere2: red $\bullet$ $M=10$, Sphere3: green $\blacksquare$, Sphere5 blue $\blacklozenge$.  }
    \label{fig:7Sp}
\end{figure}

Results for the \textsc{7Sp} model ares reported in \cref{fig:BT2D}. Here, we only show the curve for $M=2$ expansion terms. For larger $M$ the results change only slightly on the scale of the plot.
The immobile regions are initially empty (i.e. $c_{im}=0$) and the mobile region is uniformly initialised with value $c_{m}>0$. 

Notice the change in the slope at the very beginning of the curve, which is reported in more detail in the inset in \cref{fig:BT2D} for all the different expansion. This variation in due to the fact that we started from a non-equilibrium situation, where the immobile regions are completely empty and the beginning of the simulation is dominated by the exchange between mobile and immobile regions. Notice that using a different number of modes leads to different results up to $M>10$, after which the breakthrough curve does not change significantly  and the dynamic described is much faster that the one obtained with two modes only. That is because, in this case, two modes are insufficient to capture all the relevant characteristic times of the system. Conversely, ten modes are enough to capture all the relevant time scales of this system. Notice that the black line representing zero modes (i.e., no MRMT) starts decaying at later times with respect to the case where the MRMT model is employed. This can be explained as a consequence of choosing initial conditions $c_{im} = 0 \neq c_m$, which result in a net mass exchange from the mobile region to the immobile regions. As can be seen in \cref{fig:BT2D}, such mass is then released slowly at later times (a characteristic of the MRMT) at a much slower rate, since the difference in concentration between mobile and immobile regions, and consequently the net flux, is much smaller than at early times.

\begin{figure*}
  \begin{subfigure}[b]{0.5\textwidth}
   \centering
    \includegraphics[width=\textwidth]{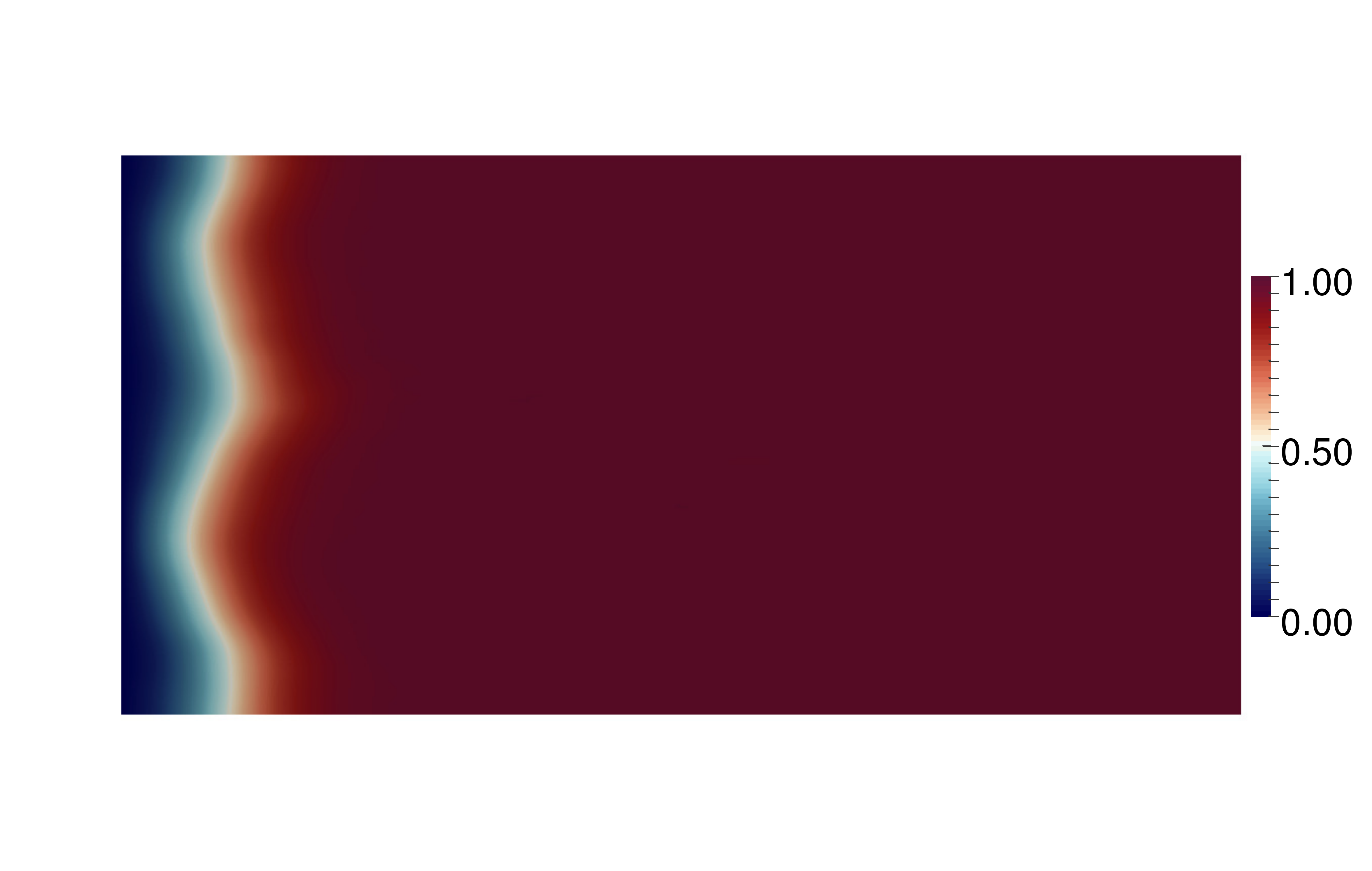}
        \vspace{-7ex}
    \caption{ t=500 s  (0.14 h) } 
  \end{subfigure}
 \begin{subfigure}[b]{0.5\textwidth}
    \centering
   \includegraphics[width=\textwidth]{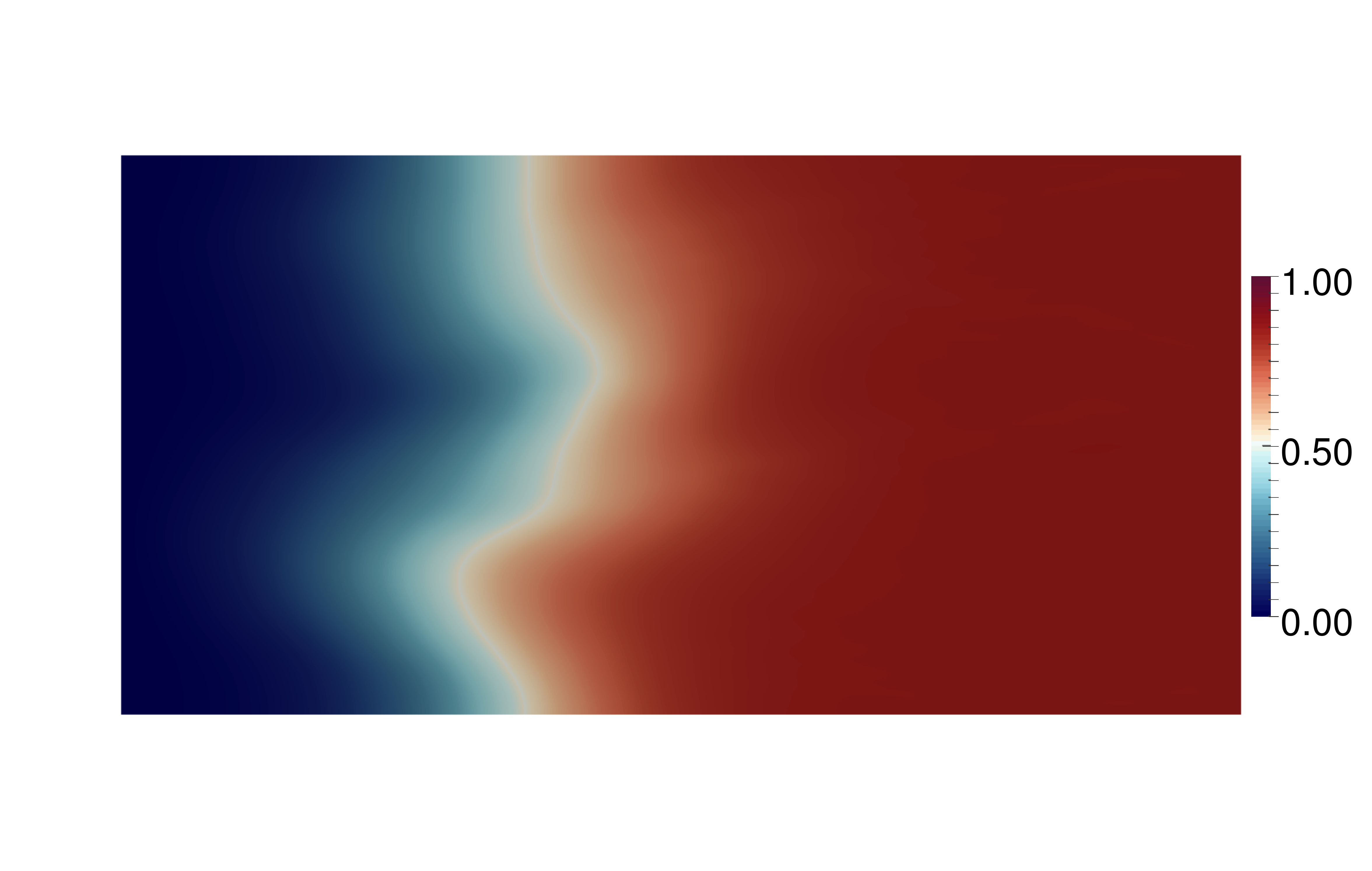}
           \vspace{-7ex}
    \caption{ t=2500 s (0.70 h)} 
  \end{subfigure} 
    \begin{subfigure}[b]{0.5\textwidth}
   \centering
    \includegraphics[width=\textwidth]{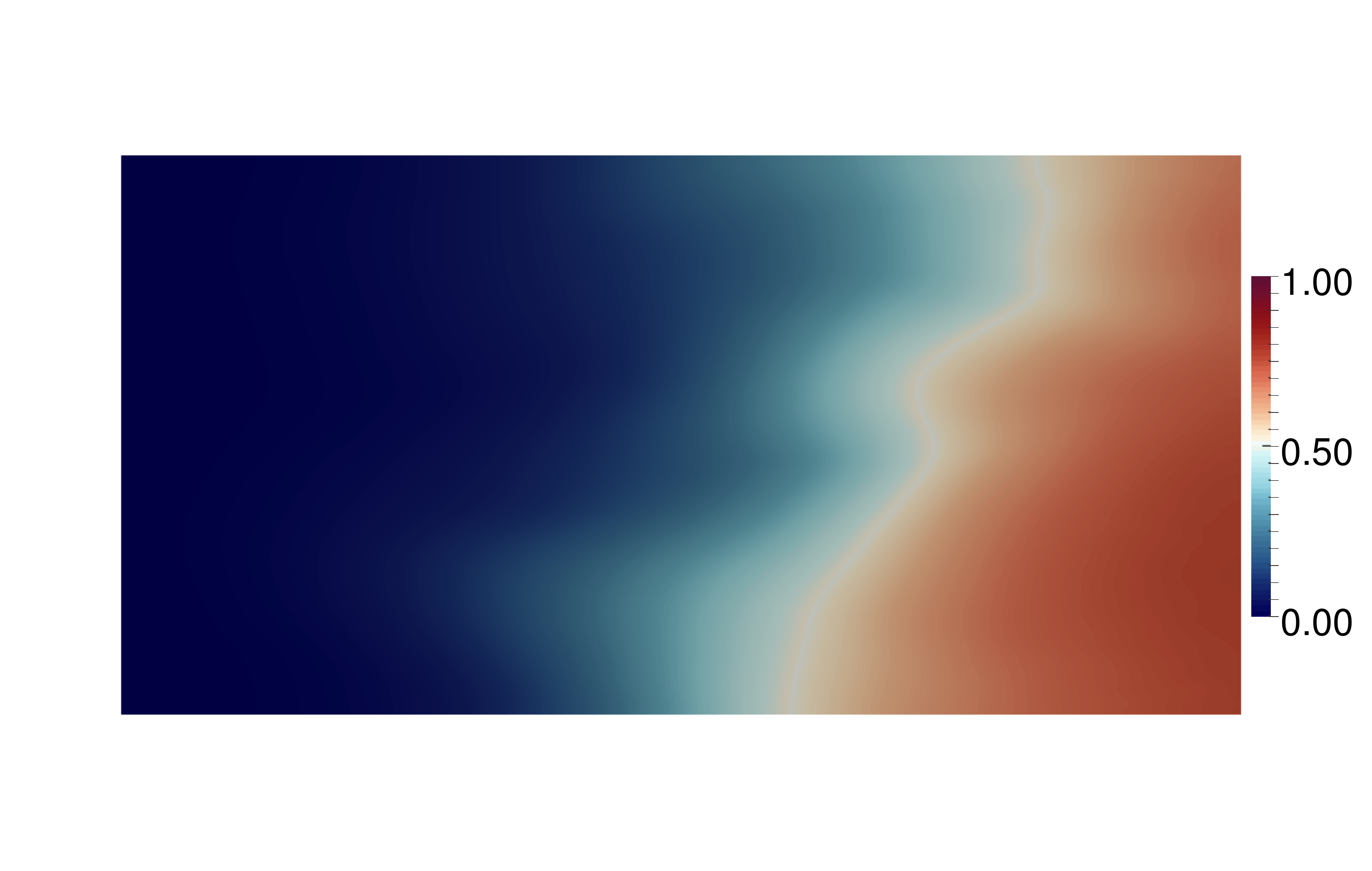}
               \vspace{-7ex}
    \caption{ t=5000 s (1.4 h)} 
  \end{subfigure}
      \begin{subfigure}[b]{0.5\textwidth}
   \centering
    \includegraphics[width=\textwidth]{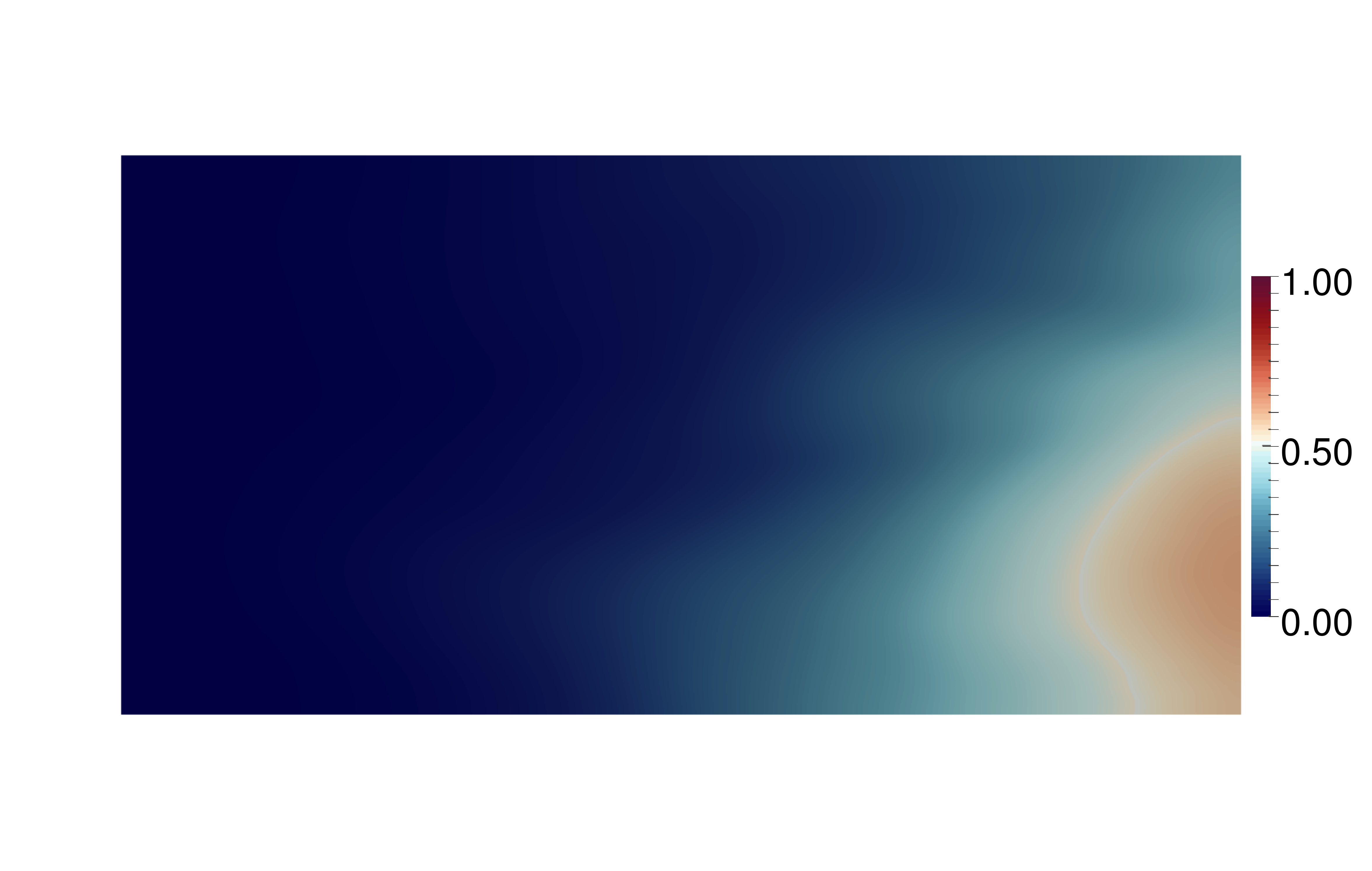}
           \vspace{-7ex}
    \caption{ t=17000 s (4.7 h)} 
  \end{subfigure}
 \caption{  Contour plot of the concentration of the concentration in the mobile region (kg/m$^3$). Results are in seconds (s) and hours (h) for the sake of readability.}
    \label{fig:2DmobCP}
\end{figure*}

\begin{figure*}
  \begin{subfigure}[b]{0.5\textwidth}
   \centering
    \includegraphics[width=\textwidth]{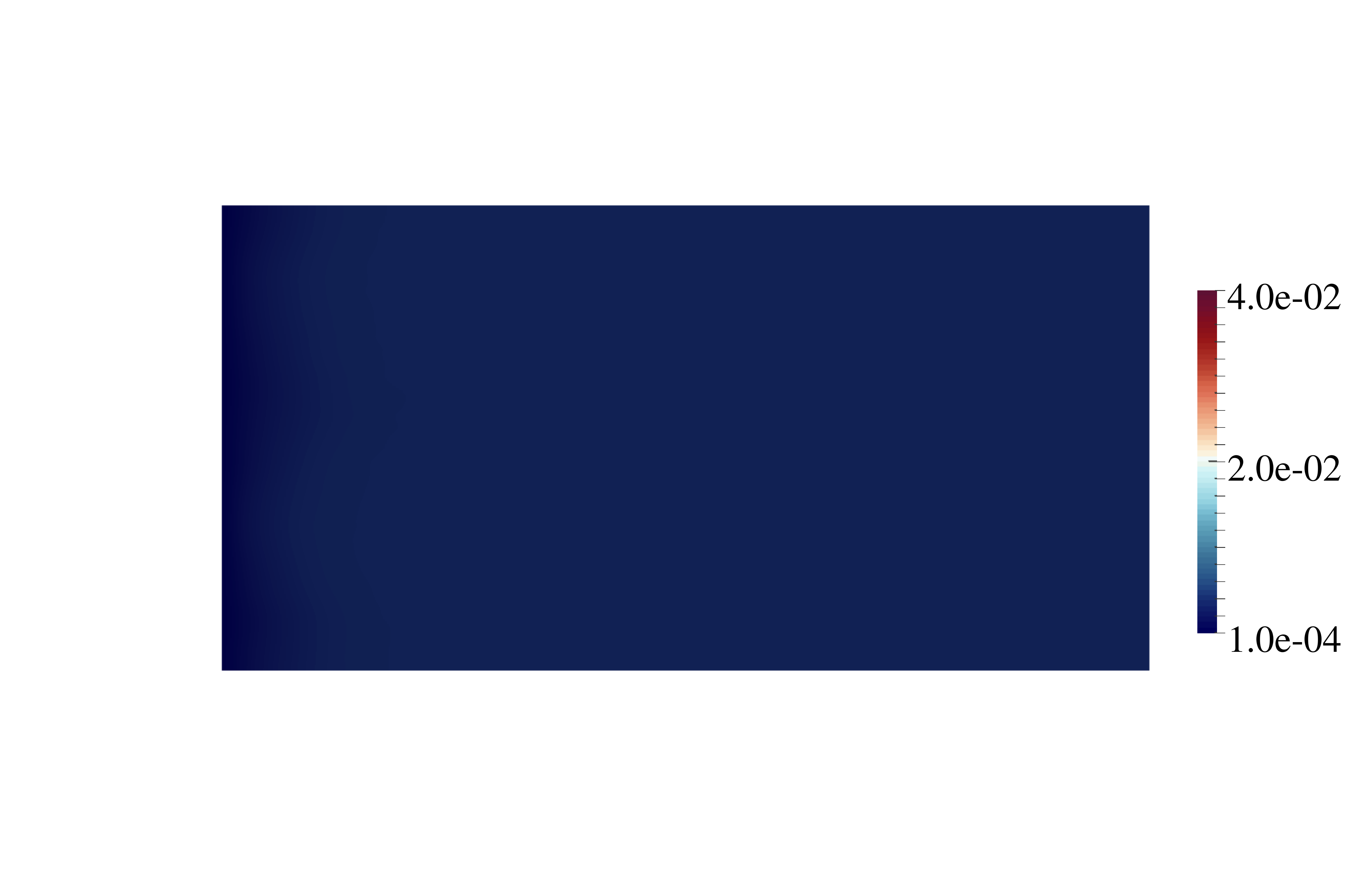}
        \vspace{-7ex}
    \caption{ t=0 s (0 h) } 
  \end{subfigure}
 \begin{subfigure}[b]{0.5\textwidth}
    \centering
   \includegraphics[width=\textwidth]{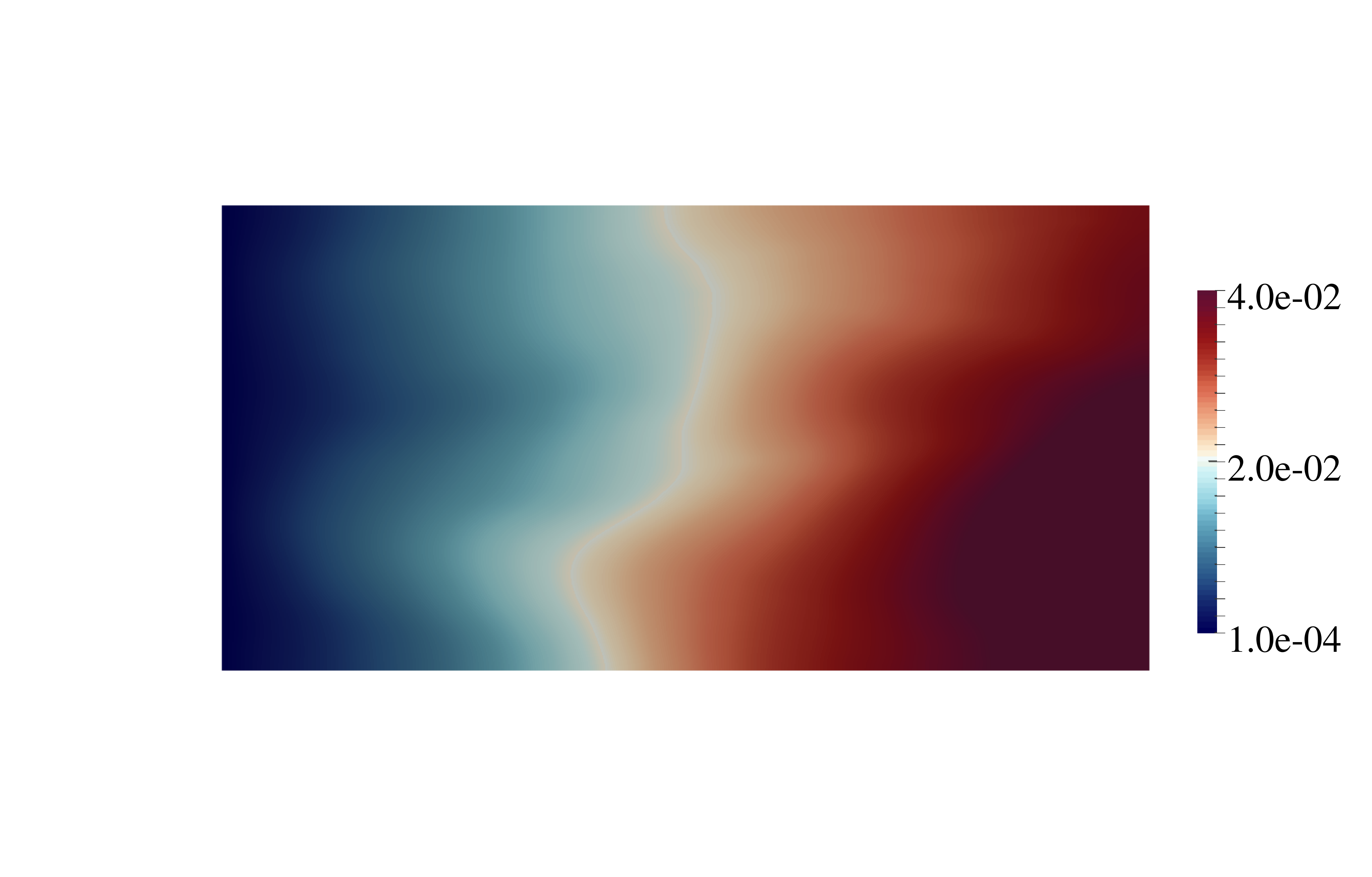}
           \vspace{-7ex}
    \caption{ t=25000 s (6.9 h) } 
  \end{subfigure} 
    \begin{subfigure}[b]{0.5\textwidth}
   \centering
    \includegraphics[width=\textwidth]{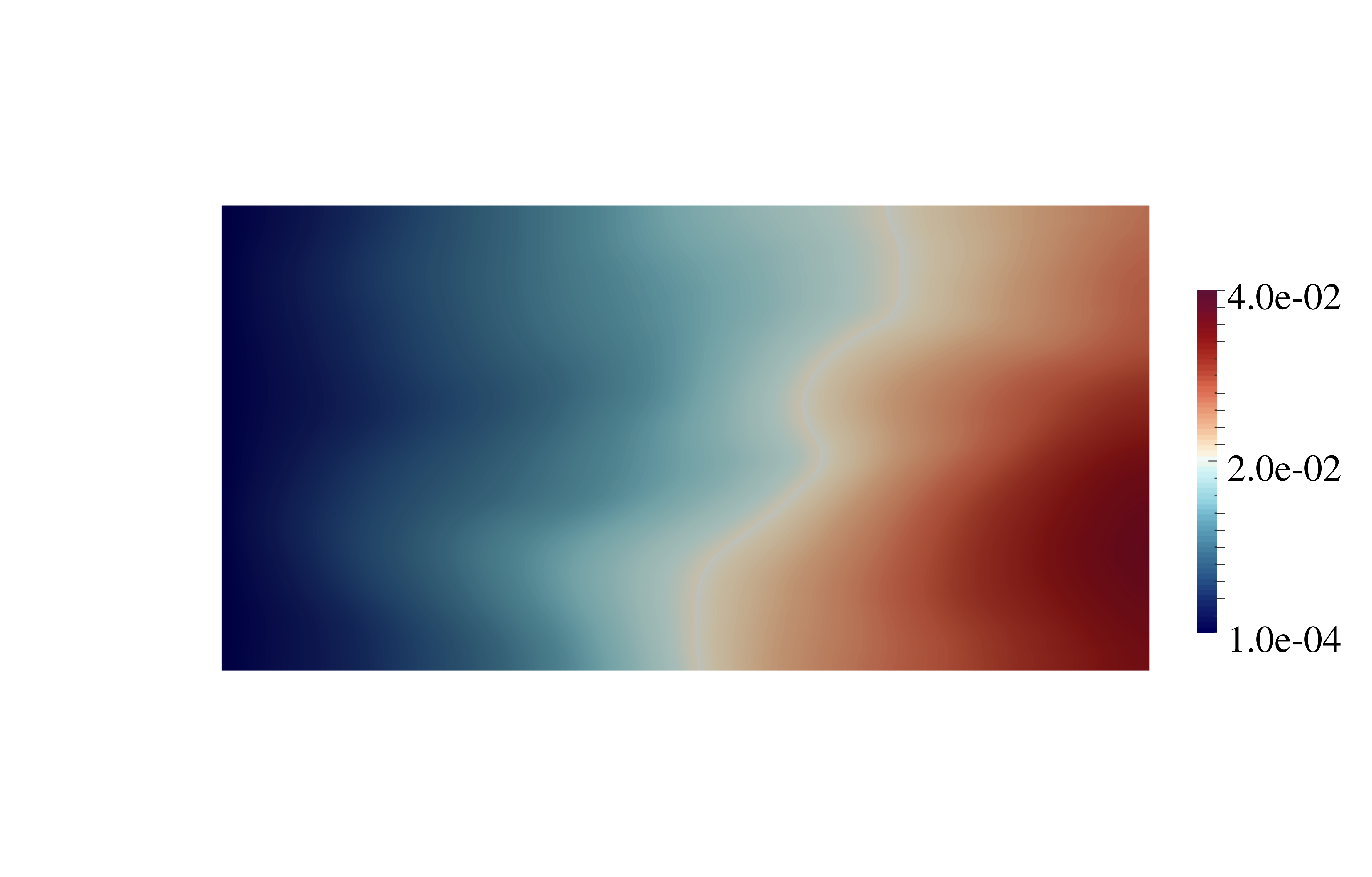}
               \vspace{-7ex}
    \caption{ t=50000 s (13.9 h)} 
  \end{subfigure}
      \begin{subfigure}[b]{0.5\textwidth}
   \centering
    \includegraphics[width=\textwidth]{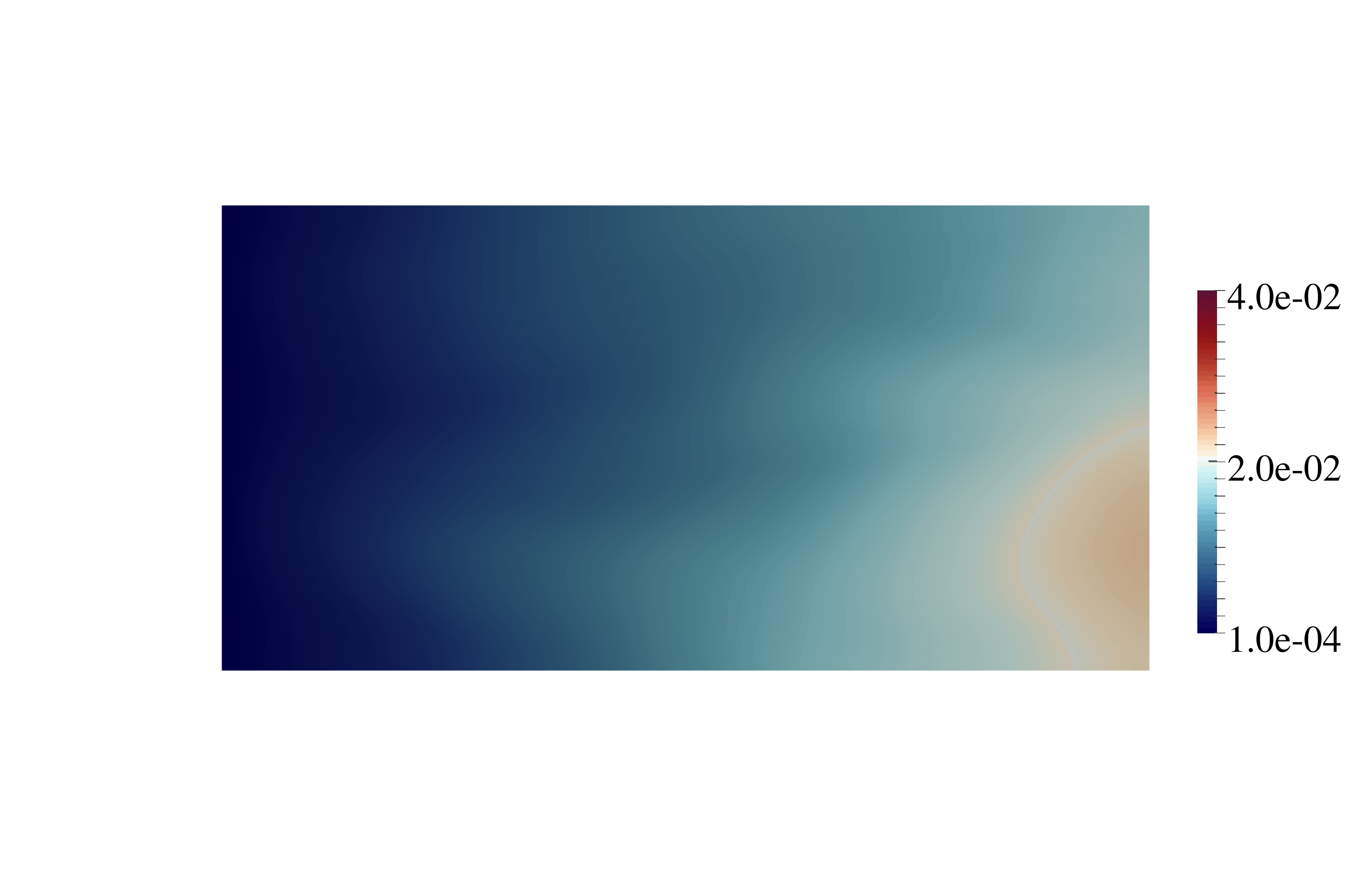}
           \vspace{-7ex}
    \caption{ t=100000 s (27.8 h) } 
  \end{subfigure}
 \caption{  Contour plot of the concentration in (kg/m$^3$) of the concentration in the immobile region (sphere1 as defined in \cref{table:7Sp}) for the Sphere1 (see \cref{table:7Sp}). Results are in seconds (s) and hours (h) for the sake of readability.}
    \label{fig:2DimmobCP}
\end{figure*}

The rapid exchange of concentration between the mobile and immobile regions is qualitatively shown in \cref{fig:2DmobCP,fig:2DimmobCP}  where the contour plots at different times of the variation of the mobile components (\cref{fig:2DmobCP}) and  and immobile one (\cref{fig:2DimmobCP}) in the 2D domain are reported. As can be observed, the process is nearly completed after approximately one day. A more quantitative result of the time variation of the concentration into the immobile region is reported in \cref{fig:7SpConc}, where the process of accumulation of the concentration following by its discharge can be observed for three different immobile regions.

\begin{figure*}
  \begin{subfigure}[b]{0.5\textwidth}
   \centering
    \includegraphics[width=0.95\textwidth]{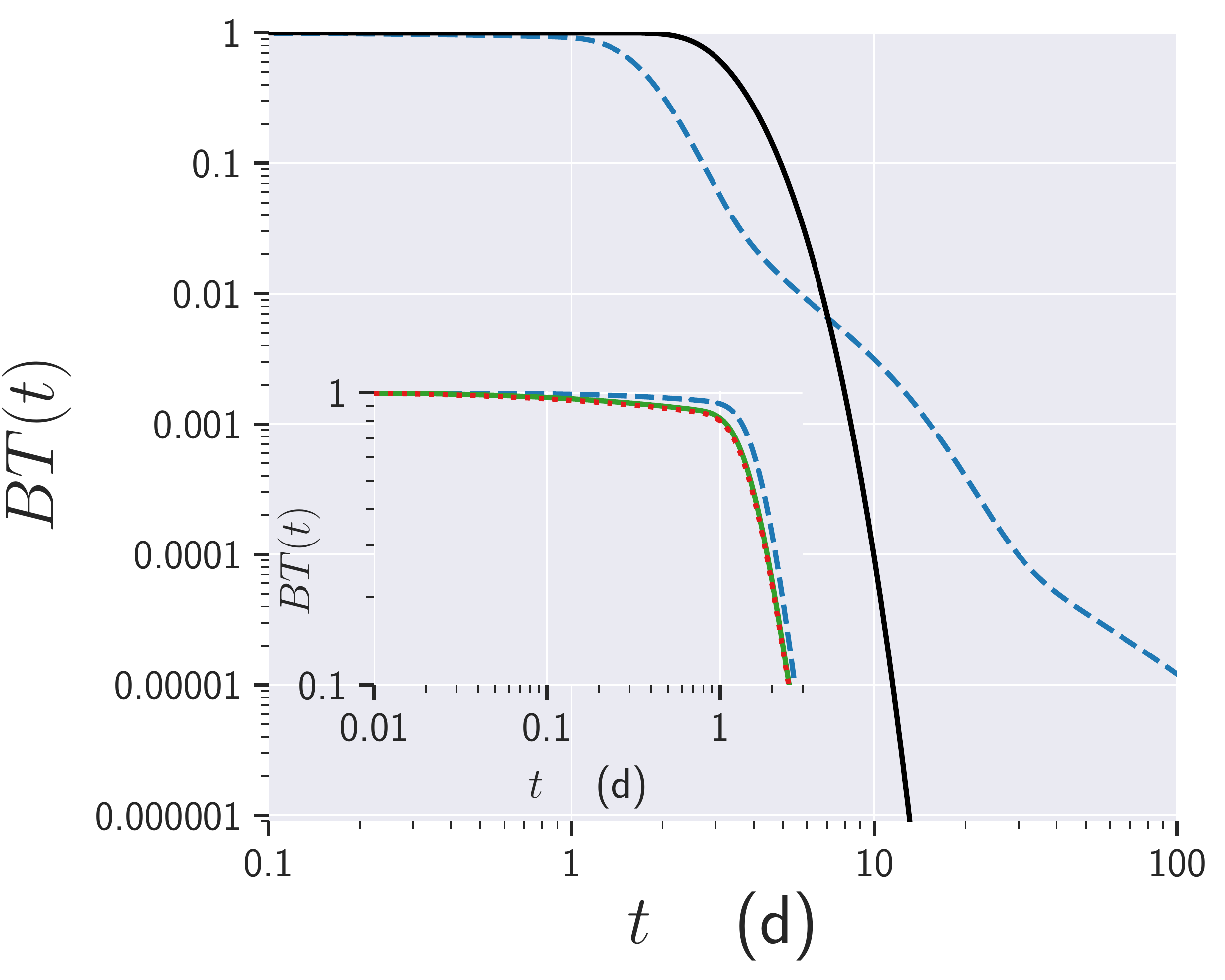}
    \caption{}
  \end{subfigure}
 \begin{subfigure}[b]{0.5\textwidth}
    \centering
   \includegraphics[width=0.9\textwidth]{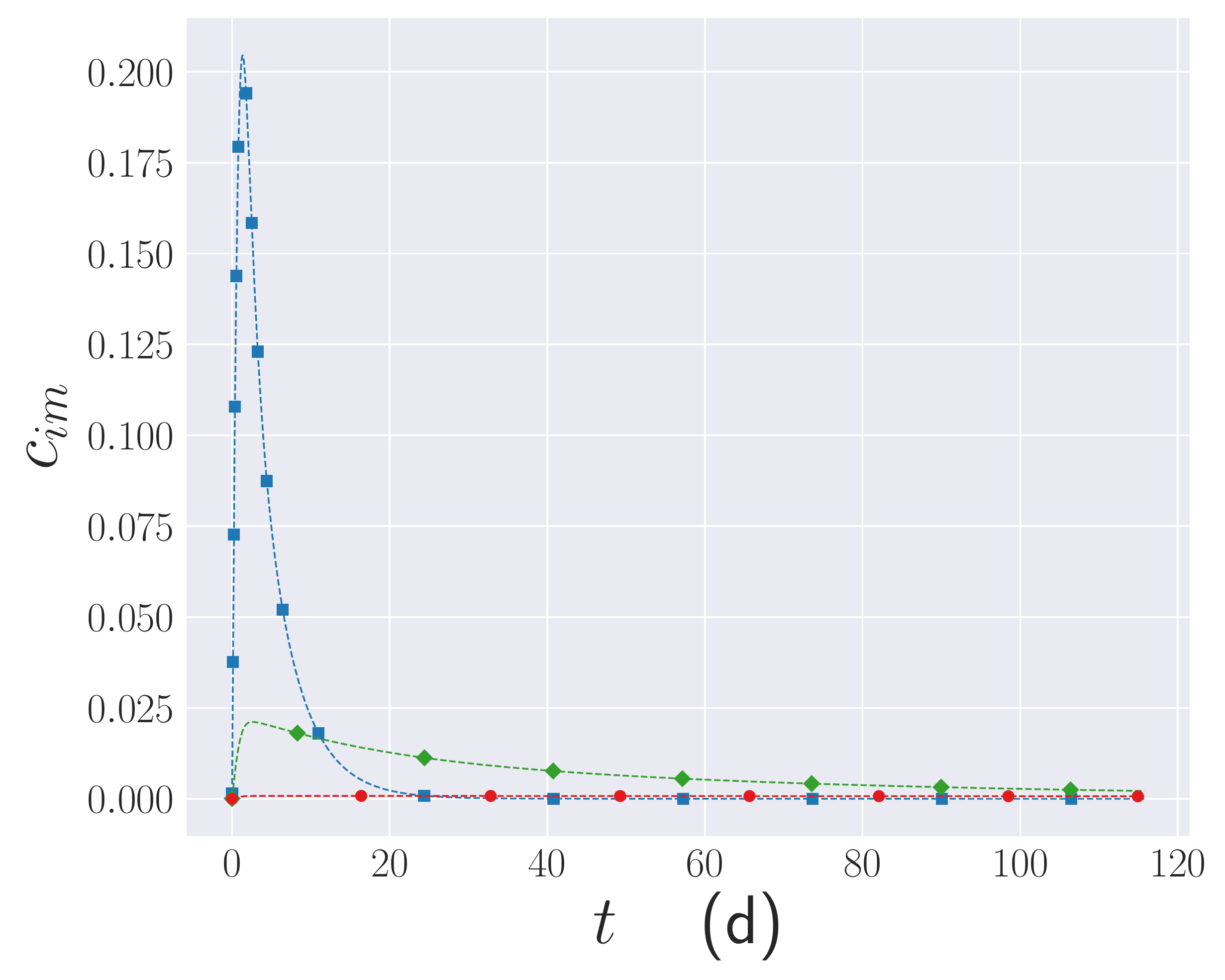}
   \caption{}
    \vspace{2ex}
  \end{subfigure} 
 \caption{ On the left: Breakthrough curves for the \textsc{Comp} case and $M=2$ and comparison with the case without multi-rate (black dashed curve). In the inset, are reported the results for the first 3 days for all the expansion considered: $M=2$ dashed blue curve, $M=10$ green continuous curve, $M=20$ dotted curve, $M=50$ dash-dotted curve. On the right: Concentration of the concentration $cm$ (in kg/m$^3$)  versus time in three of the immobile region in the \textsc{Comp} model.  Sphere2: red $\bullet$ $M=10$, Cylinder1: green $\blacksquare$, FirstOrder1 blue $\blacklozenge$ 
 }
    \label{fig:composite}
\end{figure*}

\noindent \textbf{{\textsc{Comp}} Model:} The \textsc{Comp} model gives results similar to the one presented for the \textsc{7Sp} model, and they are summarised in \cref{fig:composite}. We can still observe a transient at the beginning of the simulation, which however, results much faster than the one shown in the \textsc{7Sp} model (see \cref{fig:BT2D}).

\clearpage
\subsubsection{Heterogeneous mass transfer properties}

The last case we consider for multi-rate processes in geological media is the one in which all the relevant parameters ($K$, $\alpha$ and $\beta$) are spatially distributed. Here, we consider only one immobile region modelled as a sphere.

\begin{figure*}
  \begin{subfigure}[b]{0.5\textwidth}
   \centering
    \includegraphics[width=0.95\textwidth]{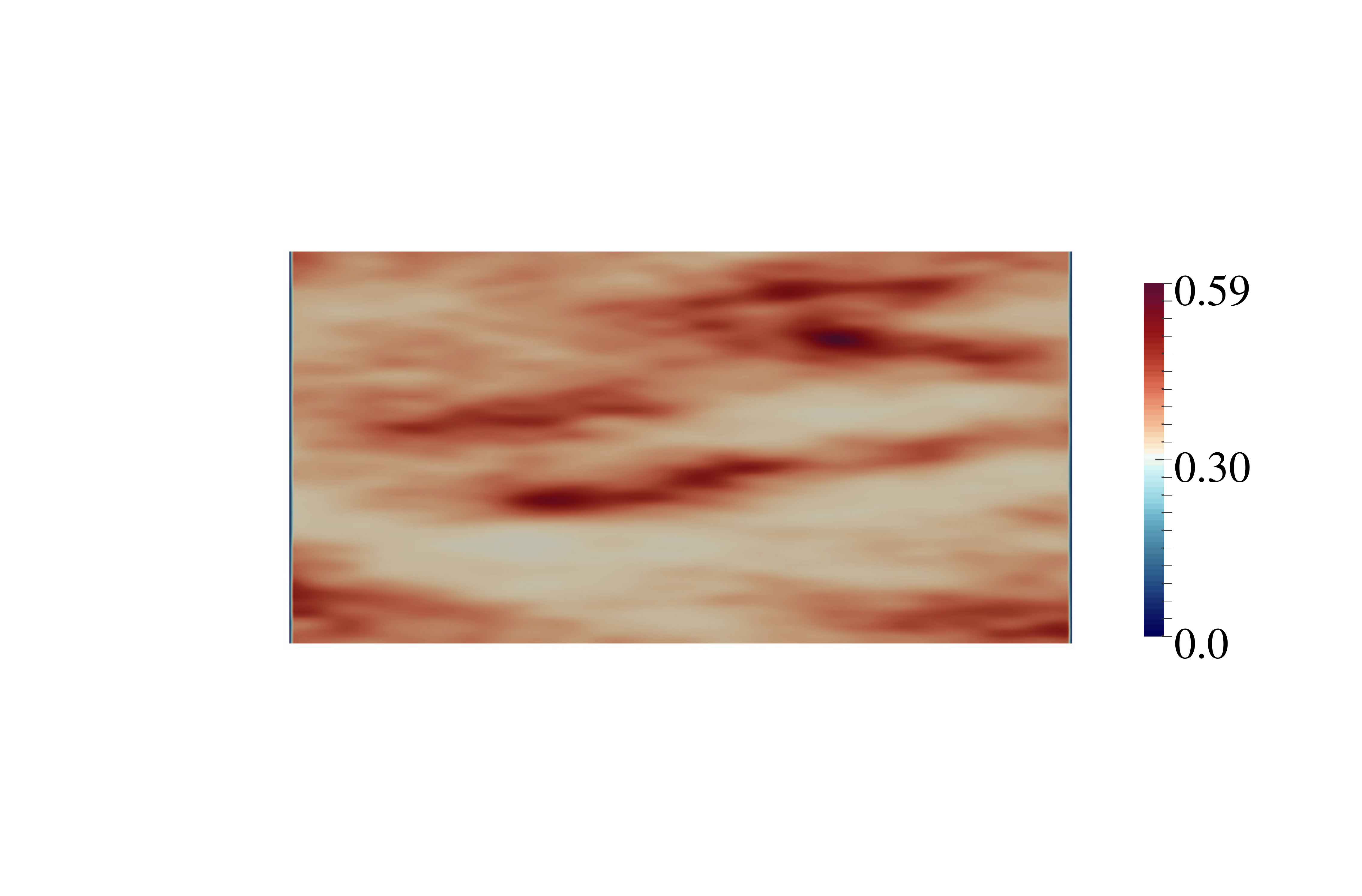}
         \vspace{-10ex}
         \caption{$\beta_m$} 
              \label{fig:Rbeta}
  \end{subfigure}
 \begin{subfigure}[b]{0.5\textwidth}
    \centering
   \includegraphics[width=0.9\textwidth]{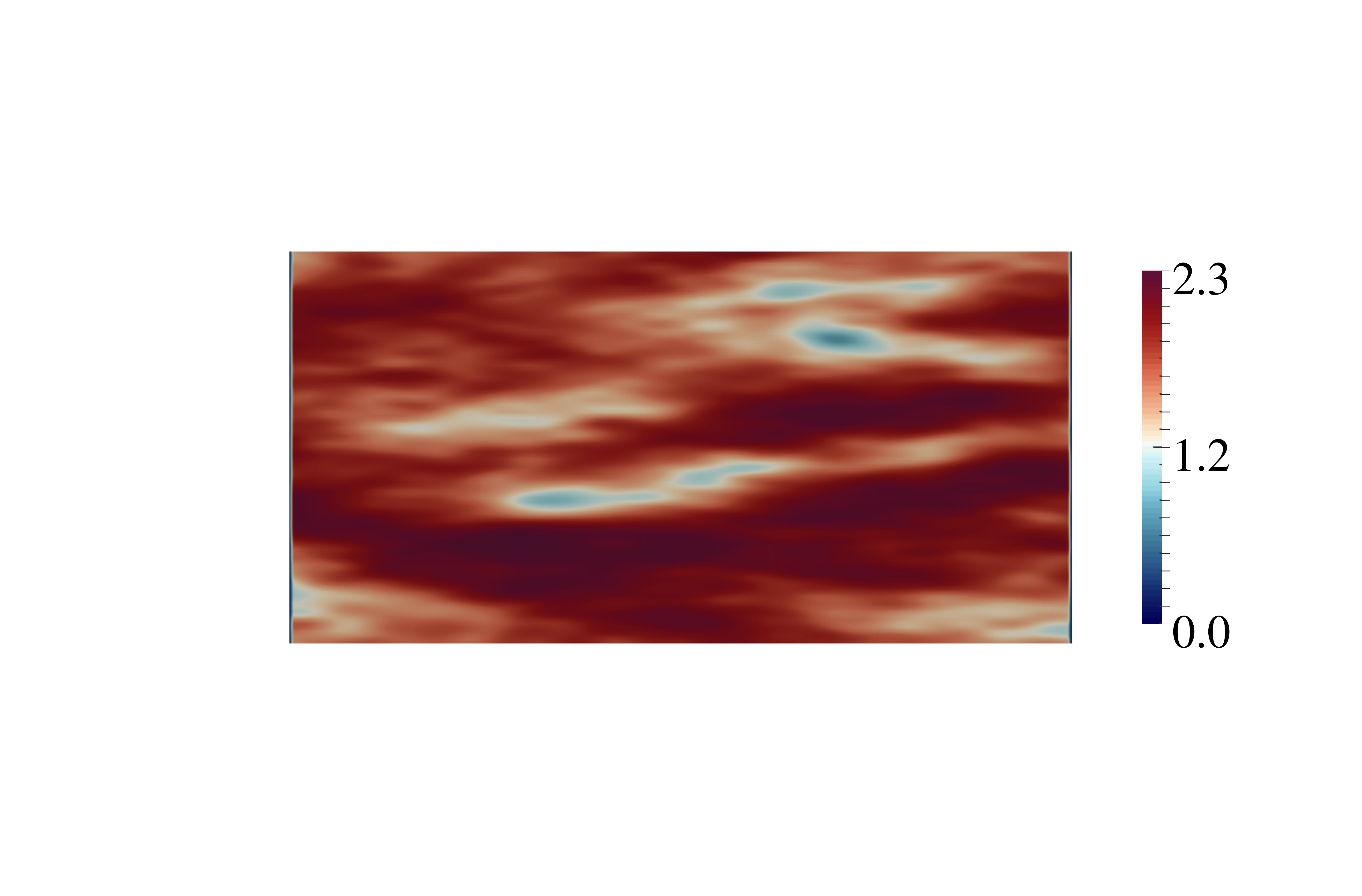}
    \vspace{-10ex}
    \caption{$\beta_{imm}$} 
        \label{fig:RbetaImm}
  \end{subfigure} 
   \begin{subfigure}[b]{\textwidth}
    \centering
   \includegraphics[width=0.6\textwidth]{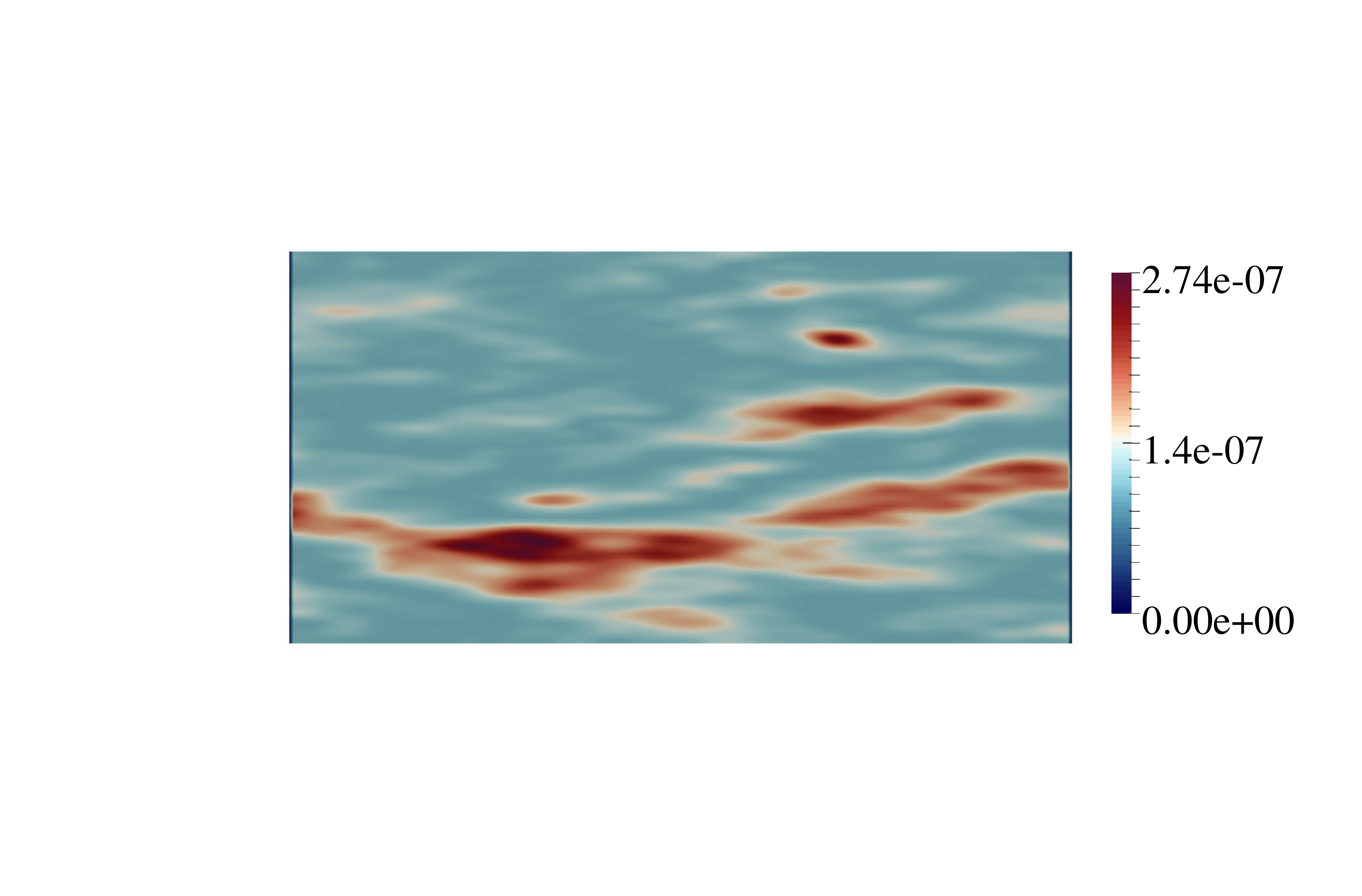}
       \vspace{-11ex}
    \caption{$\omega$} 
    \label{fig:ROm}
  \end{subfigure} 
 \caption{ Contour plot of the fields for $\beta$, $\beta_{imm}$ and $\omega$ generated using \cref{eq:randparm} }
    \label{fig:random2D}
\end{figure*}

We start by building the $\beta_m$ field which we choose to be determined by $K$. To this end,  we map the permeability random  field $K$  shown in \cref{fig:Krand} into a field for $\beta_m$ by associating to each $k$ a value in $[0.3,0.6]$. The final field for $\beta_m$ is shown in \cref{fig:Rbeta}. Then,  we obtain for $\beta_{im}$ from $\beta_m$ and $\omega$  using the Kozeny-Carman law. In the end, we employ the following expressions:
\begin{align}\label{eq:randparm} 
    \beta_m &= 0.3 + 0.6\frac{k-k_{min}}{k_{max} - k_{min}}\, , \\ \nonumber
    \omega  &= a\frac{\beta_m^3}{k(1-\beta_m)^2}\, , \\
    \beta_{im} &= \frac{1-\beta_m}{\beta_m}\, , \nonumber 
\end{align}
where $a=10^{-10}$, while $k_{max}$ and $k_{max}$ are respectively the maximum and minimum of $k$ in the domain. 
These fields are shown in \cref{fig:random2D}.

\begin{figure*}
  \begin{subfigure}[b]{0.5\textwidth}
   \centering
    \includegraphics[width=0.95\textwidth]{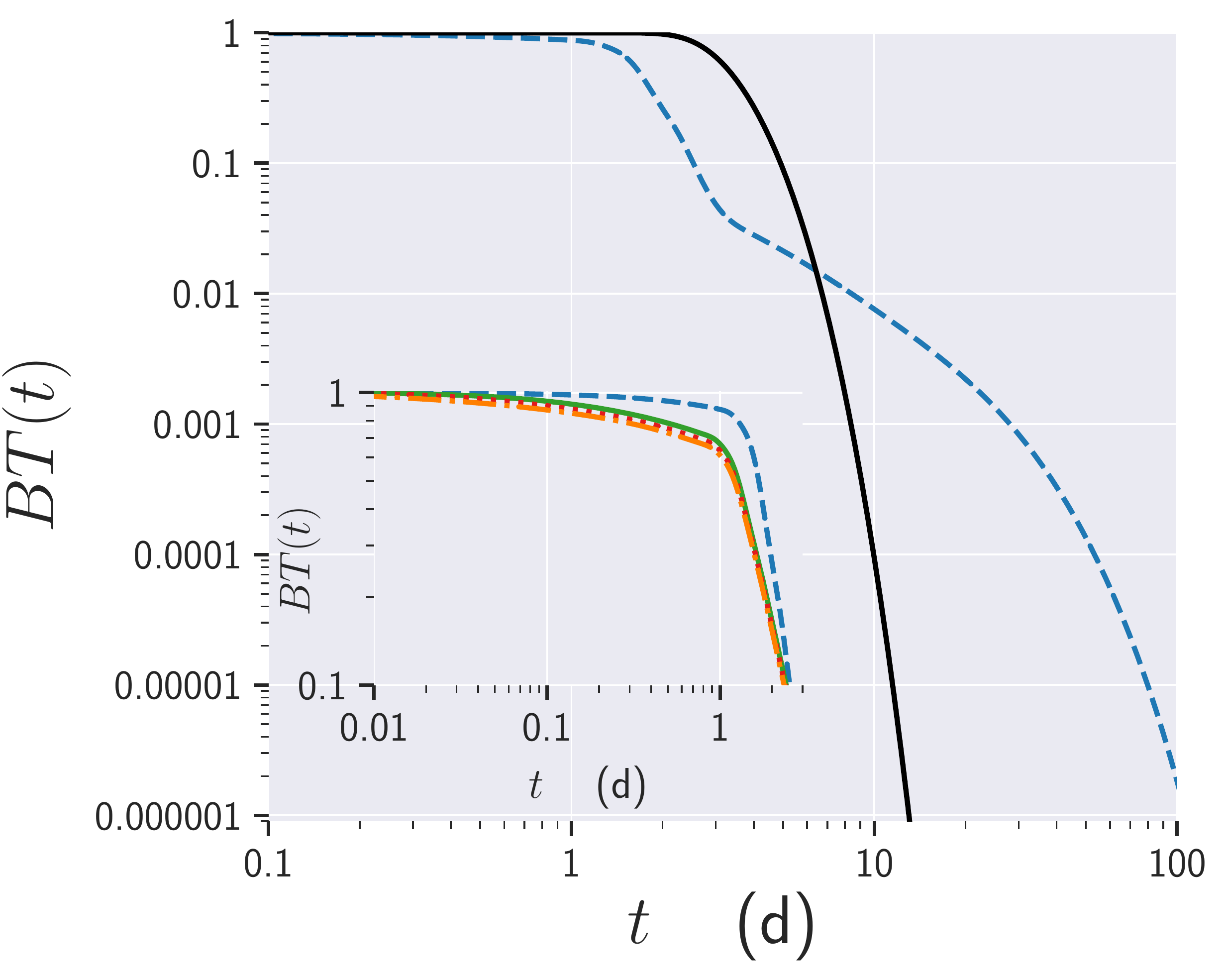}
    \caption{}
  \end{subfigure}
 \begin{subfigure}[b]{0.5\textwidth}
    \centering
   \includegraphics[width=0.9\textwidth]{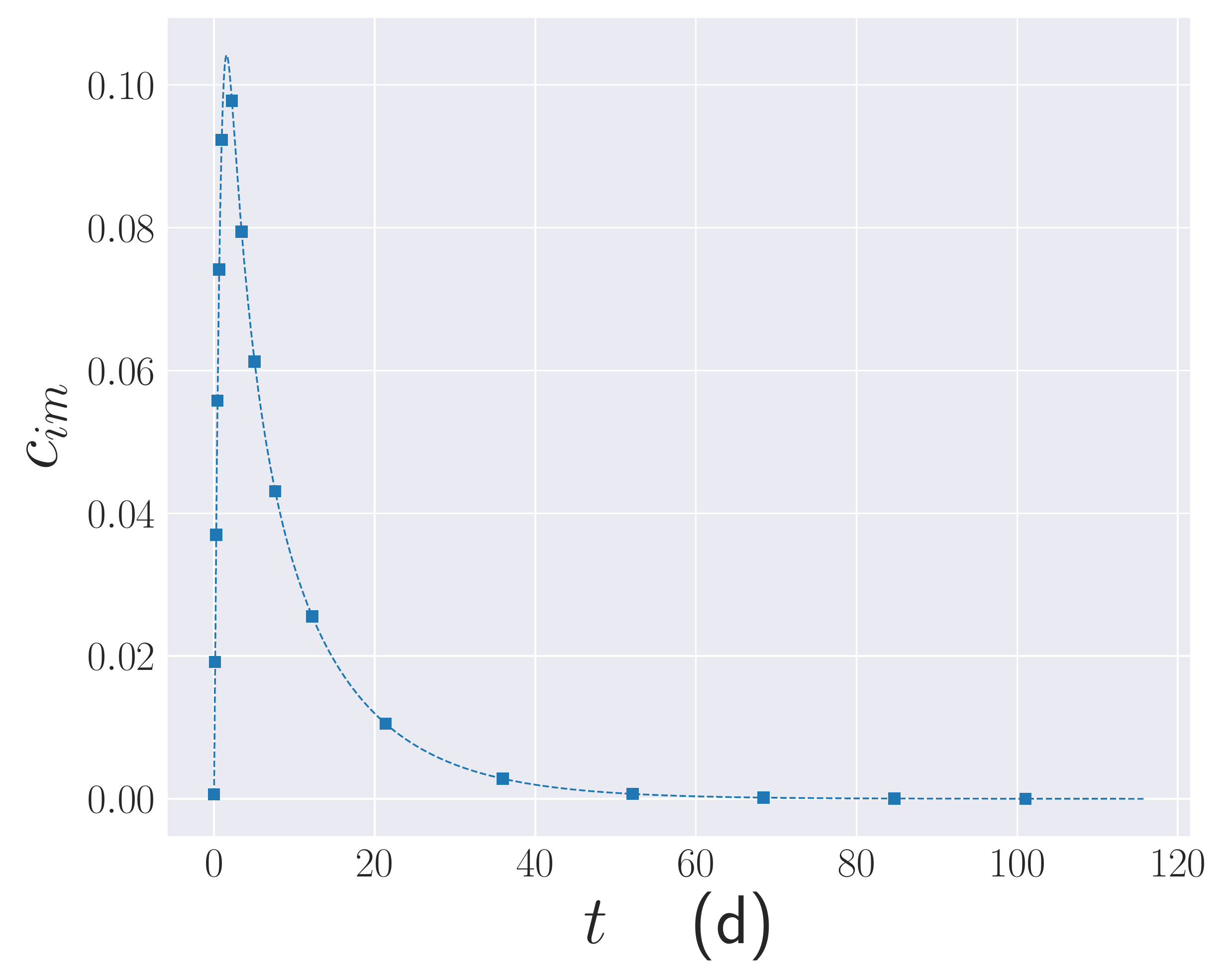}
   \caption{}
    \vspace{2ex}
  \end{subfigure} 
 \caption{ On the left: Breakthrough curves for the R\textsc{1Sp} case and $M=2$ and comparison with the case without multi-rate (black continuous curve). In the inset, are reported the results for the first 3 days for all the expansion considered: $M=2$ dashed blue curve, $M=10$ green continuous curve, $M=20$ dotted curve, $M=50$ dash-dotted curve. On the right: Concentration of the concentration $c_{im}$ (in kg/m$^3$)  versus time in the immobile region }
    \label{fig:randoms1}
\end{figure*}

Results, in terms of the concentration leaving the domain, and the time evolution of the concentration field in the immobile region are reported in \cref{fig:randoms1}. 


At early times, we can observe a transient similar to that observed in the previous cases (see \cref{fig:BT2D,fig:composite}). In this case we notice that the  peak for the average concentration in the immobile region is less pronounced than in the previous cases. However, it is interesting to notice that the breakthrough (which refers to the concentration in the mobile region) decreases faster than in the cases with homogeneous distribution.

\clearpage
\subsection{Three-dimensional packed bed}

The MRMT model in its current form does not include any chemical reaction terms. While the application to chemical engineering problems requires the modelling of chemical reactions, we are presenting here a proof of concept of the application of MRTM to a chemical engineering relevant problem, namely a packed bed column. The model is general enough to be used for these kinds of problems and working in our group in undergoing to include chemical reaction terms in the MRTM formulation.

Packed bed columns occupy a predominant role in chemical industry \citep{Jafari2008}, where they are used in several different processes such as separation, filtration, purification or as a reaction unit and their modelling at different level and scales \citep{Guo2019,Boccardo2019,Horsch2019} which can also include the modelling of the reactions inside the domains \cite{Boccardo2018}.
The flow field and transport phenomena (like heat and mass transfer) inside these equipment can be described at different level of resolution, from particle-resolved models, where the components of the packing material (such as spheres) are fully described \citep{Boccardo2015,Municchi2018,Municchi2017,Singhal2017}, to continuum heterogeneous multiphase models \citep{Municchi2018a,Cloete2018} and pseudo-homogeneous models which consider a single phase as the MRMT. Generally, bridging between different scales is not a trivial task \citep{Radl2018}.

Within the MRMT framework, the description of a packed bed with immobile particles is, on a first approximation, identical to that presented  in the previous section (see \cref{Sec:2D}) for geological media. A steady-state Darcy equation \cref{Eq:Darcy} can be solved on a randomly generated permeability field, while non-local transport can be modelled using the MRMT.

A three-dimensional cylindrical with height $0.5$ m and base diameter of $0.06$ m constituting the physical domain is discretised on a mesh composed by by 64000 cells. 
In order to achieve an accurate and bounded solution, we employed the following discretisation schemes (we direct the reader to the \textsc{OpenFOAM} user guide \cite{Openfoam2019} for a detailed description of each scheme):
\begin{itemize}
    \item divergence:      bounded Gauss vanLeer01 \cite{vanLeer1974};
    \item gradient: cellLimited  leastSquares 1;
    \item surface normal gradient: default limited 1;
\end{itemize}
These numerical schemes used ensure that the concentration of the chemical species inside the packed bed remains bounded between zero and the initial value.

Permeability and the flow field inside the domain were generated as described in the previous section by solving the steady-state Darcy equation (see \cref{Eq:Darcy}). We used the \textsc{7Sp} model described earlier (see \cref{table:7Sp}) to represent the immobile regions.

\begin{figure*}
  \begin{subfigure}[b]{0.5\textwidth}
   \centering
    \includegraphics[width=\textwidth]{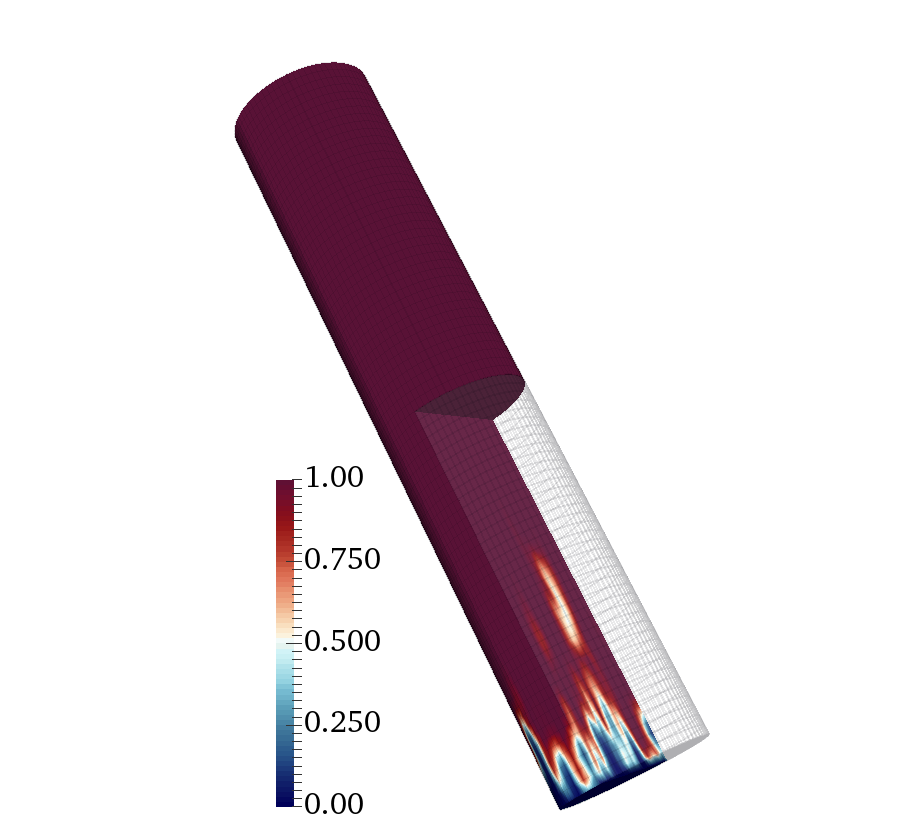}
    \caption{ t=2 s } 
  \end{subfigure}
 \begin{subfigure}[b]{0.5\textwidth}
    \centering
   \includegraphics[width=\textwidth]{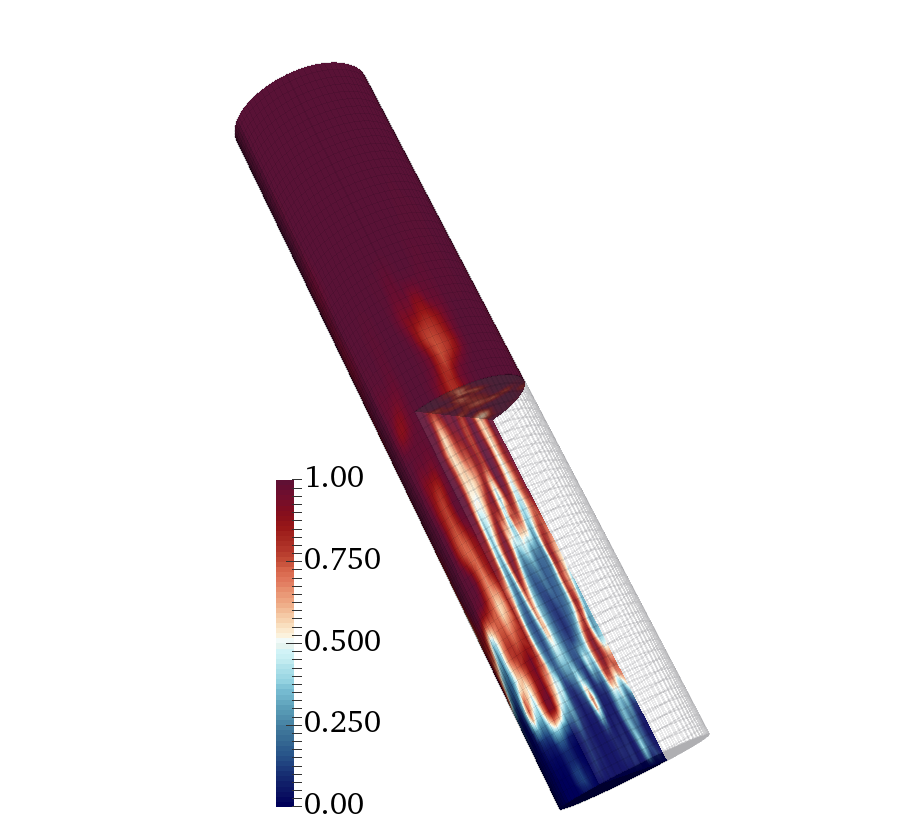}
    \caption{ t=20 s } 
  \end{subfigure} 
    \begin{subfigure}[b]{0.5\textwidth}
   \centering
    \includegraphics[width=\textwidth]{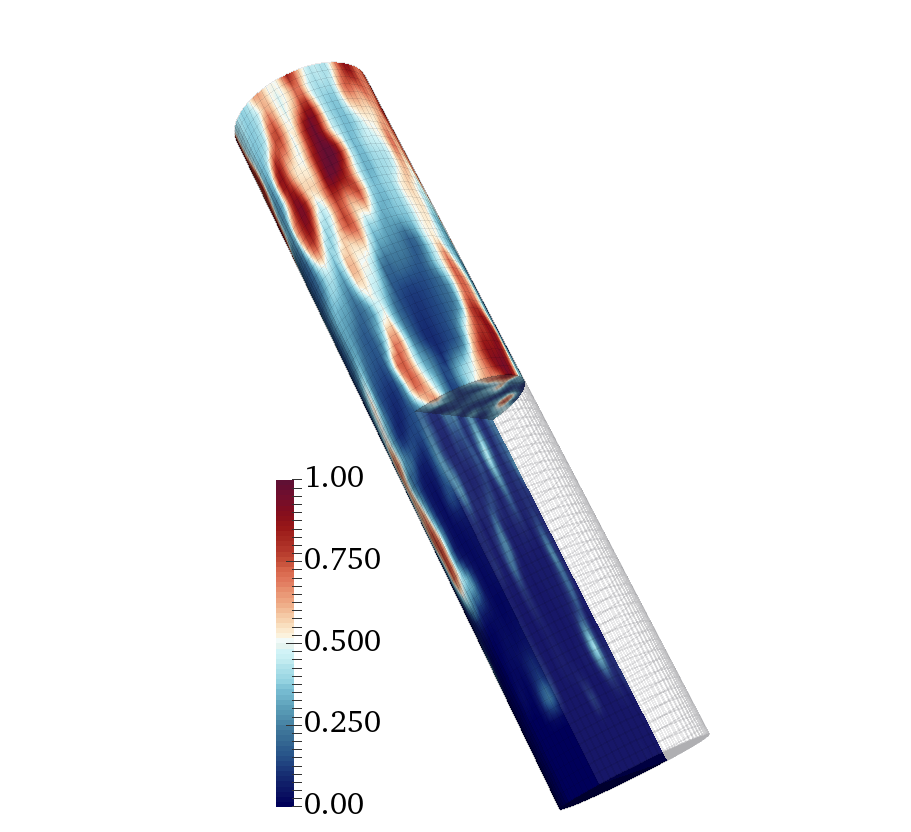}
    \caption{ t=40 s } 
  \end{subfigure}
      \begin{subfigure}[b]{0.5\textwidth}
   \centering
    \includegraphics[width=\textwidth]{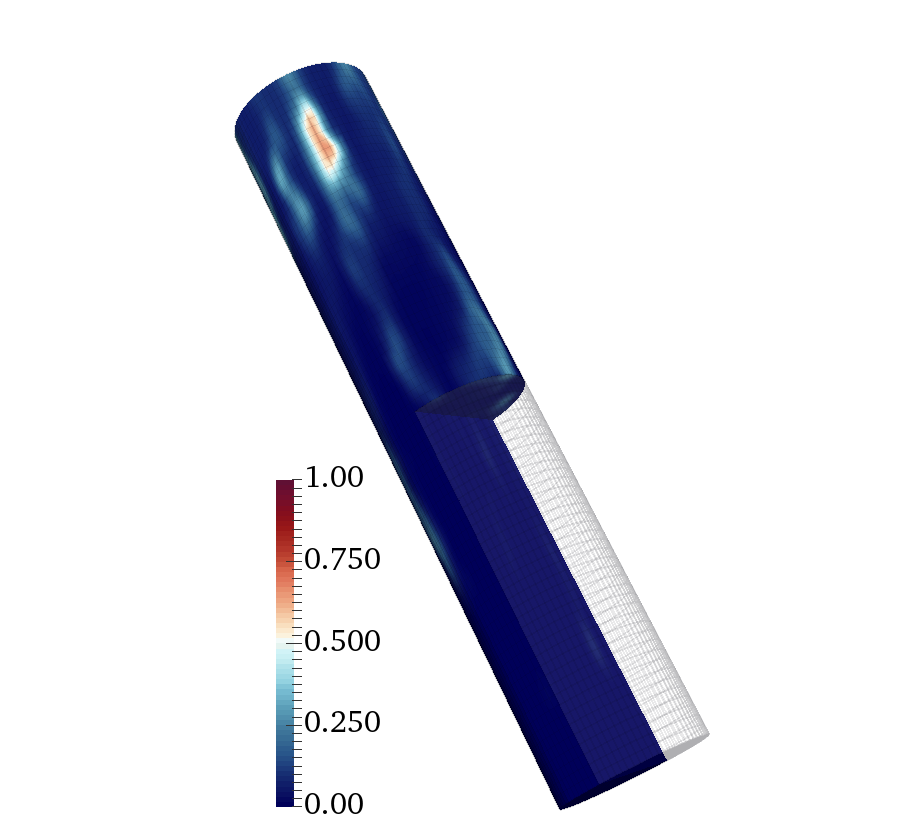}
    \caption{ t=80 s } 
  \end{subfigure}
 \caption{  Contour plot of the concentration of the chemical species in (kg/m$^3$) in the mobile region for the Packed Bed column at different times}
    \label{fig:packedB}
\end{figure*}

\begin{figure*}
   \centering
    \includegraphics[width=0.95\textwidth]{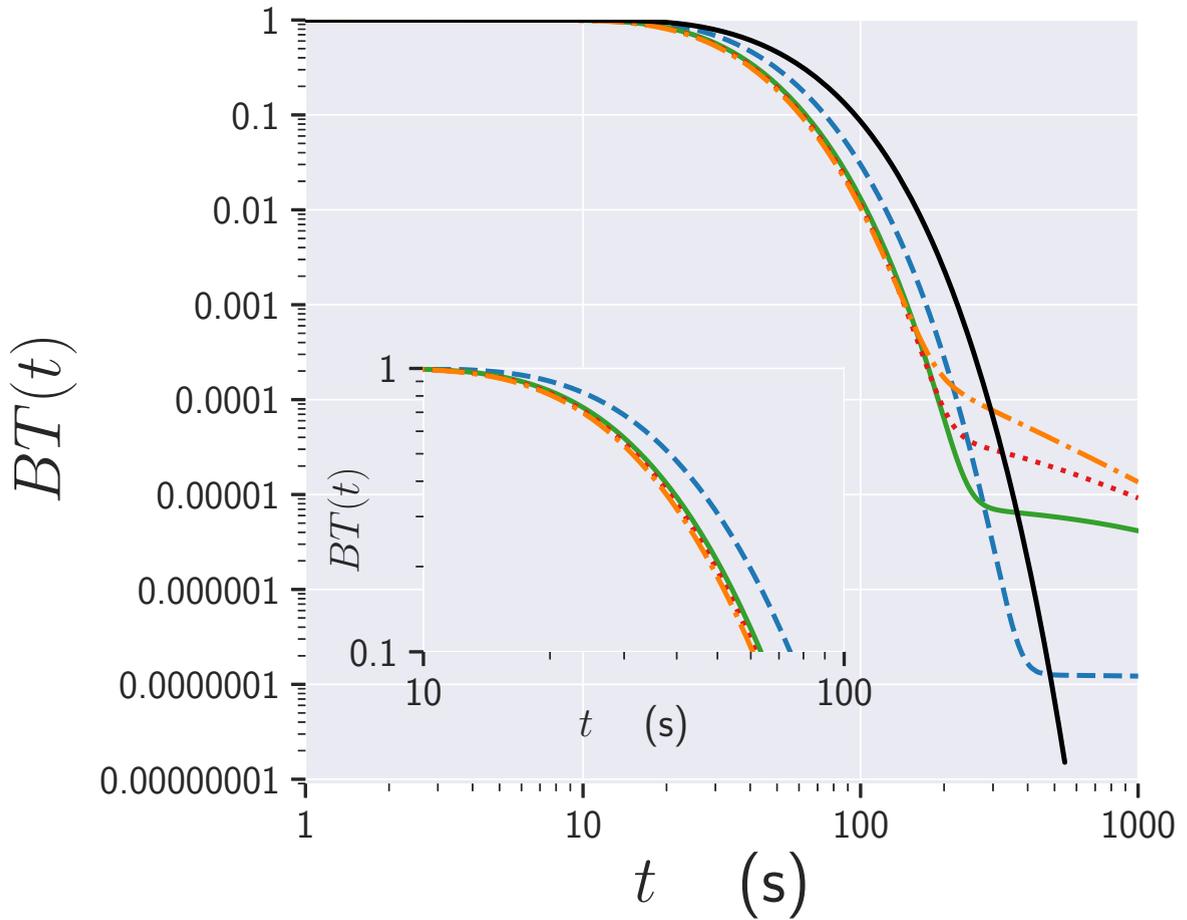}
    \label{fig:3Dbt}
 \caption{Breakthrough curves for the \textsc{7Sp} case and $M=2$ (blue dashed curve) and comparison with the the system without multi-rate (black solid curve). In the inset, are reported the results for the first 100 seconds for all the expansion considered: $M=2$ dashed blue curve, $M=10$ green continuous curve, $M=20$ red dotted curve, $M=50$ orange dash-dotted curve.   }
    \label{fig:3D7Sp}
\end{figure*}

In \cref{fig:packedB} we show the variation of the chemical species inside the column in the mobile region at different times. Notice that, as detailed in \cref{fig:3D7Sp}, there exist two different phenomena that govern mass transport inside the packed bed: (i) mass transfer between heterogeneously distributed mobile and immobile regions and, (ii) channelling due to the heterogeneous permeability field. While the second phenomenon is most prominent in the early times, the first is responsible for the long (almost horizontal) tail in \cref{fig:3D7Sp}

\FloatBarrier










\clearpage


\section{Conclusions}
In this work, we presented  a software package included in \textsc{OpenFOAM}\textsuperscript{\textregistered} to solve problems involving non-local (in time) transport phenomena using the Multi-Rate Mass Transfer model first proposed in \citfull{Haggerty1995} and subsequently formalised in \citfull{Municchi2020}. 
Such package has been developed using the  \textsc{OpenFOAM}\textsuperscript{\textregistered} library, which has a wide range of users in academia and industry alike.
The main novelties and advantages introduced in this implementation include:

\begin{itemize}
    \item Possibility to work with heterogeneous fields such as permeability, porosity, or properties of the immobile regions within the domain.
    \item The \textsc{OpenFOAM}\textsuperscript{\textregistered} technology on which this work is based allows to perform three-dimensional simulations in parallel architectures using state-of-the-art linear solvers.
    \item Being structured as an object-oriented C++ library, this software can be easily extended and integrated in other \textsc{OpenFOAM}\textsuperscript{\textregistered} solvers to perform multi-physics simulations.
\end{itemize}

We showed that the the numerical solver included in the library produces results in agreement with previous works and  calculations performed with \textsc{Chebfun} \cite{Driscoll2014}, and can reproduce results presented in literature for porous media \citep{Haggerty1995,Kaale2011}. Furthermore, we proposed a number of cases that illustrate possible applications to chemical engineering and geological media, where the method is able to capture salient features of heat/mass transport. 
Further applications could extend to ionic transport in batteries and porous media with adsorption/desorption reactions or flow in fractures with stagnation zones. 

\section*{Acknowledgements}
This work has been funded by the European Union's Horizon 2020
research and innovation programme, grant agreement number 764531, "SECURe -- Subsurface Evaluation of Carbon capture and storage and Unconventional risks".

%
%

\bibliographystyle{cas-model2-names}

\bibliography{bibliography_20Feb20}
\end{document}